\documentclass[showpacs,preprintnumbers,amsmath,amssymb]{revtex4}
\usepackage{graphicx}% Include figure files
\usepackage{dcolumn}% Align table columns on decimal point
\usepackage{bm}% bold math

\begin{document}

\title{
k-ESSENCE, AVOIDANCE OF THE WEINBERG'S\\ COSMOLOGICAL CONSTANT
NO-GO THEOREM AND OTHER\\  DARK ENERGY EFFECTS OF  TWO MEASURES
FIELD THEORY}

\author
{E. I. Guendelman \thanks{guendel@bgu.ac.il} and A.  B. Kaganovich
\thanks{alexk@bgu.ac.il}}
\address{Physics Department, Ben Gurion University of the Negev, Beer
Sheva 84105, Israel}

\date{\today}% It is always \today, today,

\begin{abstract}
The dilaton-gravity sector of the Two Measures Field Theory (TMT)
is explored in detail in the context of cosmology.  The model
possesses scale invariance which is spontaneously broken due to
the intrinsic features of the TMT dynamics. The dilaton $\phi$
dependence of the effective Lagrangian appears only as a result of
the spontaneous breakdown of the scale invariance. If no fine
tuning is made, the effective $\phi$-Lagrangian $p(\phi,X)$
depends quadratically upon the kinetic energy $X$. Hence TMT may
represent an explicit example of {\it the effective} $k$-{\it
essence resulting from first principles without any exotic term}
in the fundamental action intended for obtaining this result.
Depending of the choice of regions in the parameter space, TMT
exhibits different possible outputs for cosmological dynamics: a)
Possibility of a power law inflation driven by the field $\phi$
which is followed by the late time evolution driven both by a
small cosmological constant and the field $\phi$ with a
quintessence-like potential. TMT enables two ways for achieving
small cosmological constant without fine tuning of dimensionfull
parameters: either by a {\it seesaw} type mechanism or due to
 {\it a correspondence principle} between TMT and
conventional field theories (i.e theories with only the measure of
integration $\sqrt{-g}$  in the action). b) Possibility of {\it
resolution of the old cosmological constant problem}. From the
point of view of TMT, it becomes clear why the old cosmological
constant problem cannot be solved (without fine tuning) in
conventional field theories. c) The power law inflation without
any fine tuning can end with damped oscillations of $\phi$ around
the state with zero cosmological constant. d) There is a broad
range of the parameters such that: the equation-of-state in the
late time universe $w=p/\rho <-1$; $w$ {\it asymptotically} (as
$t\rightarrow\infty$) {\it approaches} $-1$ {\it from below};
$\rho$ approaches a cosmological constant, the smallness of which
does not require fine tuning of dimensionfull parameters.
\end{abstract}

   \pacs{98.80.Cq, 04.20.Cv, 95.36.+x}% PACS, the Physics and Astronomy
\maketitle

\section{Introduction}

The cosmological constant (CC) problem \cite{Weinberg1}-\cite{CC},
the accelerated expansion of the late time universe\cite{accel},
the cosmic coincidence \cite{coinc} are challenges for the
foundations of modern physics (see also reviews on dark
energy\cite{de-review}-\cite{Copeland}, dark matter
\cite{dm-review} and references therein). Numerous models have
been proposed with the aim to find answer to these puzzles, for
example: the quintessence\cite{quint},  coupled
quintessence\cite{Amendola}, $k$-essence\cite{k-essence}, variable
mass particles\cite{vamp}, interacting quintessence\cite{int-q},
Chaplygin gas\cite{Chapl}, phantom field\cite{phantom}, tachyon
matter cosmology\cite{tachyon}, brane world scenarios\cite{brane},
  etc.. These
puzzles have also motivated an interest in modifications and even
radical revisions of the standard gravitational theory (General
Relativity (GR))\cite{modified-gravity1},\cite{modified-gravity2}.
Although motivations for most of these models can be found in
fundamental theories like for example in brane
world\cite{extra-dim},  the questions concerning the Einstein GR
limit and relation to the regular particle physics, like standard
model, still remain unclear.

It is very hard to imagine that  it is possible  to propose ideas
 which are able to solve  the above mentioned problems  keeping
 at the same time unchanged the basis of fundamental physics, i.e.
 gravity and particle field theory. This paper may be regarded as
 {\it an attempt to find a way for satisfactory answers at least to a part of the
 fundamental questions on the basis of first principles}, i.e. without
using semi-empirical models. In this paper we explore a toy model
including gravity and a single scalar field $\phi$ in the
framework of the so called Two Measures Field Theory
(TMT)\cite{GK1}-\cite{GK8}. In TMT, gravity (or more exactly,
geometry) and particle field theory intertwined in a very non
trivial manner, but the Einstein's GR is restored when the fermion
matter energy density is much larger than the vacuum energy
density\cite{GK6},\cite{GK7}.

Here we {\it have no  purpose} of constructing a complete
realistic cosmological scenario. Instead, we concentrate our
attention on the possible role TMT may play in resolution of a
number of the above mentioned problems. The field theory model we
will study here is invariant under global scale transformation of
the metric accompanied  with an appropriate shift of the dilaton
field $\phi$. This scale symmetry is spontaneously broken due to
intrinsic features of the TMT dynamics. Except for a peculiar
structure of the TMT action, the latter does not contain any
exotic terms. The obtained dynamics represents an explicit example
of $k$-essence resulting from first principles.

 The parameter space of the model
allows to find different regions where the following different
effects can take place {\it without fine tuning of the parameters
and initial conditions}:

a) Possibility of a power law inflation driven by a scalar field
$\phi$ which is  followed by the late time evolution  driven both
by a small cosmological constant and a scalar field $\phi$ with a
quintessence-like potential; smallness of the cosmological
constant can be achieved {\it without fine tuning of dimensionfull
parameters}.

b) In another region of the parameters, there is a possibility of
the power law inflation  ended with damped oscillations of $\phi$
around the state with zero cosmological constant (we want to
emphasize again that this is realized without fine tuning of the
parameters and initial conditions). Thus this scenario includes
{\it an exit from inflation together with  resolution of the old
CC problem}. This becomes possible because  the basic assumptions
of the Weinberg's no-go theorem\cite{Weinberg1} are generically
violated in TMT models.

c) There is a region of the parameters where the model exhibits
the possibility for a superacceleration phase of the universe: in
the late time universe the dark energy density $\rho$ increases
asymptotically (as $t\rightarrow\infty$) approaching from below to
a constant value; the equation-of-state $p/\rho =w <-1$ and $w$
{\it asymptotically approaches} $-1$ {\it from below}.

The organization of the paper is the following: In Sec.II we
present a  review of the basic ideas of TMT. Sec.III starts from
formulation of a simple scale invariant model containing all the
terms respecting the symmetry of the model but without any exotic
terms. In Appendix A we present equations of motion in the
original frame. Using results of Appendix A, the complete set of
equations of motion in the Einstein frame is given in Sec.III as
well. It is shown there that if no fine tuning of the parameters
is made, the effective action of our model in the Einstein frame
turns out to be a k-essence type action. We start in Sec.IV from a
simple case with fine tuned parameters where the non-linear
dependence of the kinetic term disappears. Then three different
shapes of the effective potential are possible. For the spatially
flat FRW universe we study some features of the cosmological
dynamics for each of the shapes of the effective potential. For
one of the shapes of the effective potential, the zero vacuum
energy is achieved without any fine tuning.  In Sec.V we study the
cosmological dynamics without tuning parameters. It is shown that
there is the possibility of a superacceleration phase of the
universe, and some details of the dynamics are explored. Sec.VI is
devoted to studying in more details the CC problems: a) we discuss
two mechanisms for resolution of the new CC problem; b) we analyze
the difference between TMT and conventional field theories (where
only the measure of integration $\sqrt{-g}$ is used in the
fundamental action) which allows to understand why from the point
of view of TMT the conventional field theories failed to solve the
old CC problem. In Appendix B we shortly discuss what kind of
model one would obtain when choosing fine tuned couplings to the
measures of integration in the action. Some additional remarks
concerning the relation between the structure of TMT action and
the CC problem are given in Appendix C.

\section{Basis of Two Measures Field Theory}

TMT is a generally coordinate invariant theory where {\it all the
difference from the standard field theory in curved space-time
consists only of the following three additional assumptions}:

1. The main supposition  is that for describing the effective
action for 'gravity $+$ matter' at energies below the Planck
scale, the usual form of the action $S = \int L\sqrt{-g}d^{4}x$ is
not complete. Our positive hypothesis is that the effective action
has to be of the form\cite{GK3}-\cite{GK8}
\begin{equation}
    S = \int L_{1}\Phi d^{4}x +\int L_{2}\sqrt{-g}d^{4}x
\label{S}
\end{equation}
 including two Lagrangians $ L_{1}$ and $L_{2}$ and two
measures of integration $\sqrt{-g}$ and $\Phi$ or, equivalently,
two volume elements $\Phi d^{4}x$ and $\sqrt{-g}d^{4}x$
respectively. One is the usual measure of integration $\sqrt{-g}$
in the 4-dimensional space-time manifold equipped with the metric
 $g_{\mu\nu}$.
Another  is the new measure of integration $\Phi$ in the same
4-dimensional space-time manifold. The measure  $\Phi$ being  a
scalar density and a total derivative\footnote{For applications of
the measure $\Phi$ in string and brane theories see
Ref.\cite{Mstring}.} may be defined
\begin{itemize}

\item
either by means of  four scalar fields $\varphi_{a}$
($a=1,2,3,4$), (compare with the approach by Wilczek\footnote{See:
 F. Wilczek, Phys.Rev.Lett. {\bf 80}, 4851
(1998). Wilczek's goal was to avoid the use of a fundamental
metric, and for this purpose he needs five scalar fields. In our
case we keep the standard role of the metric from the beginning,
but enrich the theory with a new metric independent density.}),
\begin{equation}
\Phi
=\varepsilon^{\mu\nu\alpha\beta}\varepsilon_{abcd}\partial_{\mu}\varphi_{a}
\partial_{\nu}\varphi_{b}\partial_{\alpha}\varphi_{c}
\partial_{\beta}\varphi_{d}.
\label{Phi}
\end{equation}

\item
or by means of a totally antisymmetric three index field
$A_{\alpha\beta\gamma}$
\begin{equation}
\Phi
=\varepsilon^{\mu\nu\alpha\beta}\partial_{\mu}A_{\nu\alpha\beta}.
\label{Aabg}
\end{equation}
\end{itemize}

To provide parity conservation in the case given by Eq.(\ref{Phi})
one can choose for example one of $\varphi_{a}$'s to be a
pseudoscalar; in the case given by Eq.(\ref{Aabg}) we must choose
$A_{\alpha\beta\gamma}$ to have negative parity. Some ideas
concerning the nature of the measure fields $\varphi_{a}$ are
discussed in Ref.\cite{GK8}. A special case of the structure
(\ref{S}) with definition (\ref{Aabg}) has been recently discussed
in Ref.\cite{hodge} in applications to supersymmetric theory and
the CC problem.

2. It is assumed that the Lagrangian densities $ L_{1}$ and
$L_{2}$ are functions of all matter fields, the dilaton field, the
metric, the connection
 but not of the
"measure fields" ($\varphi_{a}$ or $A_{\alpha\beta\gamma}$). In
such a case, i.e. when the measure fields  enter in the theory
only via the measure $\Phi$,
  the action (\ref{S}) possesses
an infinite dimensional symmetry. In the case given by
Eq.(\ref{Phi}) these symmetry transformations have the form
$\varphi_{a}\rightarrow\varphi_{a}+f_{a}(L_{1})$, where
$f_{a}(L_{1})$ are arbitrary functions of  $L_{1}$ (see details in
Ref.\cite{GK3}); in the case given by Eq.(\ref{Aabg}) they read
$A_{\alpha\beta\gamma}\rightarrow
A_{\alpha\beta\gamma}+\varepsilon_{\mu\alpha\beta\gamma}
f^{\mu}(L_{1})$ where $f^{\mu}(L_{1})$ are four arbitrary
functions of $L_{1}$ and $\varepsilon_{\mu\alpha\beta\gamma}$ is
numerically the same as $\varepsilon^{\mu\alpha\beta\gamma}$. One
can hope that this symmetry should prevent emergence of a measure
fields dependence in $ L_{1}$ and $L_{2}$ after quantum effects
are taken into account.

3. Important feature of TMT that is responsible for many
interesting and desirable results of the field theory models
studied so far\cite{GK3}-\cite{GK8}
 consists of the assumption that all fields, including
also metric, connection  and the {\it measure fields}
($\varphi_{a}$ or $A_{\alpha\beta\gamma}$) are independent
dynamical variables. All the relations between them are results of
equations of motion.  In particular, the independence of the
metric and the connection means that we proceed in the first order
formalism and the relation between connection and metric is not
necessarily according to Riemannian geometry.

We want to stress again that except for the listed three
assumptions we do not make any changes as compared with principles
of the standard field theory in curved space-time. In other words,
all the freedom in constructing different models in the framework
of TMT consists of the choice of the concrete matter content and
the Lagrangians $ L_{1}$ and $L_{2}$ that is quite similar to the
standard field theory.

Since $\Phi$ is a total derivative, a shift of $L_{1}$ by a
constant, $L_{1}\rightarrow L_{1}+const$, has no effect on the
equations of motion. Similar shift of $L_{2}$ would lead to the
change of the constant part of the Lagrangian coupled to the
volume element $\sqrt{-g}d^{4}x $. In the standard GR, this
constant term is the cosmological constant. However in TMT the
relation between the constant
 term of $L_{2}$ and the physical cosmological constant is very non
trivial (see \cite{GK3}-\cite{K},\cite{GK5}-\cite{GK7}).

In the case of the definition of $\Phi$ by means of
Eq.(\ref{Phi}), varying the measure fields $\varphi_{a}$, we get
\begin{equation}
B^{\mu}_{a}\partial_{\mu}L_{1}=0  \quad \text{where} \quad
B^{\mu}_{a}=\varepsilon^{\mu\nu\alpha\beta}\varepsilon_{abcd}
\partial_{\nu}\varphi_{b}\partial_{\alpha}\varphi_{c}
\partial_{\beta}\varphi_{d}.\label{varphiB}
\end{equation}
Since $Det (B^{\mu}_{a}) = \frac{4^{-4}}{4!}\Phi^{3}$ it follows
that if $\Phi\neq 0$,
\begin{equation}
 L_{1}=sM^{4} =const
\label{varphi}
\end{equation}
where $s=\pm 1$ and $M$ is a constant of integration with the
dimension of mass. In what follows we make the choice $s=1$.

In the case of the definition (\ref{Aabg}), variation of
$A_{\alpha\beta\gamma}$ yields
\begin{equation}
\varepsilon^{\mu\nu\alpha\beta}\partial_{\mu}L_{1}=0, \label{AdL1}
\end{equation}
that implies Eq.(\ref{varphi}) without the  condition $\Phi\neq 0$
needed in the model with four scalar fields $\varphi_{a}$.

 One should notice
 {\it the very important differences of
TMT from scalar-tensor theories with nonminimal coupling}: \\
 a) In general, the Lagrangian density $L_{1}$ (coupled to the measure
$\Phi$) may contain not only the scalar curvature term (or more
general gravity term) but also all possible matter fields terms.
This means that TMT modifies in general both the gravitational
sector  and the matter sector; b) If the field $\Phi$ were the
fundamental (non composite) one then instead of (\ref{varphi}),
the variation of $\Phi$ would result in the equation $L_{1}=0$ and
therefore the dimensionfull integration constant $M^4$ would not
appear in the theory.

Applying the Palatini formalism in TMT one can show (see for
example \cite{GK3} and Appendix A of the present paper)  that in
addition to the Christoffel's connection coefficients, the
resulting relation between connection and metric includes also the
gradient of the ratio of the two measures
\begin{equation}
\zeta \equiv\frac{\Phi}{\sqrt{-g}} \label{zeta}
\end{equation}
which is a scalar field. Consequently geometry of the space-time
with the metric $g_{\mu\nu}$ is non-Riemannian. The gravity and
matter field equations obtained by means of the first order
formalism contain both $\zeta$ and its gradient as well. It turns
out that at least at the classical level, the measure fields
affect the theory only through the scalar field $\zeta$.

The consistency condition of equations of motion has the form of a
constraint which determines $\zeta (x)$ as a function of matter
fields. The surprising feature of the theory is
 that neither Newton constant nor curvature appear in this constraint
which means that the {\it geometrical scalar field} $\zeta (x)$
{\it is determined by the matter fields configuration}  locally
and straightforward (that is without gravitational interaction).

By an appropriate change of the dynamical variables which includes
a conformal transformation of the metric, one can formulate the
theory in a Riemannian  space-time. The corresponding conformal
frame we call "the Einstein frame". The big advantage of TMT is
that in the very wide class of models, {\it the gravity and all
matter fields equations of motion take canonical GR form in the
Einstein frame}.
 All the novelty of TMT in the Einstein frame as compared
with the standard GR is revealed only
 in an unusual structure of the scalar fields
effective potential (produced in the Einstein frame), masses of
fermions  and their interactions with scalar fields as well as in
the unusual structure of fermion contributions to the
energy-momentum tensor: all these quantities appear to be $\zeta$
dependent\cite{GK5}-\cite{GK7}. This is why the scalar field
$\zeta (x)$ determined by the constraint as a function of matter
fields, has a key role in dynamics of TMT models.

\section{Scale invariant model}
\subsection{Symmetries and Action}

The TMT models possessing a global scale
invariance\cite{G1}-\cite{GKatz},\cite{GK5}-\cite{GK7} are of
significant interest because they demonstrate the possibility of
spontaneous breakdown of the scale symmetry\footnote{The field
theory models with explicitly broken scale symmetry and their
application to the quintessential inflation type  cosmological
scenarios have been studied in Ref.\cite{K}. Inflation and
transition to slowly accelerated phase from higher curvature terms
was studied in Ref.\cite{GKatz}. }. In fact, if the action
(\ref{S}) is scale invariant then this classical field theory
effect results from Eq.(\ref{varphi}), namely  from solving the
equation of motion (\ref{varphiB}) or (\ref{AdL1}).
 One of the
interesting applications of the scale invariant TMT
models\cite{GK5} is a possibility to generate the exponential
potential for the scalar field $\phi$ by means of the mentioned
spontaneous symmetry breaking even  without introducing any
potentials for  $\phi$ in the Lagrangians  $ L_{1}$ and $L_{2}$ of
the action (\ref{S}). Some cosmological applications of this
effect have been also studied in
 Ref.\cite{GK5}.

A dilaton field $\phi$ allows to realize a spontaneously broken
global scale invariance\cite{G1} and together with this it governs
the evolution of the universe on different stages: in the early
universe $\phi$ plays the role of inflaton and in the late time
universe it is transformed into a part of the dark energy.

According to the general prescriptions of TMT, we have to start
from studying the self-consistent system of gravity (metric
$g_{\mu\nu}$ and connection $\Gamma^{\mu}_{\alpha\beta}$), the
measure $\Phi$ degrees of freedom (i.e. measure fields $\varphi_a$
or $A_{\alpha\beta\gamma}$) and the dilaton field $\phi$
proceeding in the first order formalism.  We postulate that the
theory is invariant under the global scale transformations:
\begin{equation}
    g_{\mu\nu}\rightarrow e^{\theta }g_{\mu\nu}, \quad
\Gamma^{\mu}_{\alpha\beta}\rightarrow \Gamma^{\mu}_{\alpha\beta},
\quad \varphi_{a}\rightarrow \lambda_{ab}\varphi_{b}\quad
\text{where} \quad \det(\lambda_{ab})=e^{2\theta}, \quad
\phi\rightarrow \phi-\frac{M_{p}}{\alpha}\theta . \label{st}
\end{equation}
If the definition (\ref{Aabg}) is used for the measure $\Phi$ then
 the transformations of $\varphi_{a}$ in
(\ref{st}) should be changed by $A_{\alpha\beta\gamma}\rightarrow
e^{2\theta}A_{\alpha\beta\gamma}$. This global scale invariance
includes the shift symmetry\footnote{Compare the way the shift
symmetry is realized here with that of Ref.\cite{Carroll}.} of the
dilaton $\phi$ and this is the main factor why it is important for
cosmological applications of the
theory\cite{G1}-\cite{GKatz},\cite{GK5}-\cite{GK7}.

We choose an action which, except for the modification of the
general structure caused by the basic assumptions of TMT,
 {\it does not contain
 any exotic terms and  fields} as like in the conventional formulation
 of the minimally coupled scalar-gravity system.
Keeping the general structure (\ref{S}), it is convenient to
represent the action in the following form:
\begin{eqnarray}
&S=&\int d^{4}x e^{\alpha\phi /M_{p}}
\left[-\frac{1}{\kappa}R(\Gamma ,g)(\Phi +b_{g}\sqrt{-g})+(\Phi
+b_{\phi}\sqrt{-g})\frac{1}{2}g^{\mu\nu}\phi_{,\mu}\phi_{,\nu}-e^{\alpha\phi
/M_{p}}\left(\Phi V_{1} +\sqrt{-g}V_{2}\right)\right]
 \label{totaction}
\end{eqnarray}

In the action (\ref{totaction}) there are two types of the
gravitational terms and
 of the "kinetic-like terms"  which
respect the scale invariance : the terms of the one type coupled
to the
 measure $\Phi$ and those of the other type
coupled to the measure $\sqrt{-g}$. Using the freedom in
normalization of the measure fields ($\varphi_{a}$  in the case of
using Eq.(\ref{Phi}) or $A_{\alpha\beta\gamma}$ when using
Eq.(\ref{Aabg})), we set the coupling constant of the scalar
curvature to the measure $\Phi$ to be  $-\frac{1}{\kappa}$.
Normalizing all the fields such that their couplings to the
measure $\Phi$ have no additional factors, we are not able in
general to provide the same in terms describing the appropriate
couplings to the measure $\sqrt{-g}$. This fact explains the need
to introduce the dimensionless real parameters $b_g$ and
$b_{\phi}$ and we will only assume that {\it they have close
orders of magnitudes}. Note that in the case of the choice
$b_g=b_{\phi}$ we would proceed with {\it the fine tuned model}.
The real positive parameter $\alpha$ is assumed to be of the order
of unity; in all solutions presented in this paper we set $\alpha
=0.2$. For Newton constant we use $\kappa =16\pi G $, \,
$M_p=(8\pi G)^{-1/2}$.

One should  also point out the possibility of introducing two
different pre-potentials which are exponential functions of the
dilaton $\phi$ coupled to the measures $\Phi$ and $\sqrt{-g}$ with
factors $V_{1}$ and $V_{2}$. Such $\phi$-dependence provides the
scale symmetry (\ref{st}). However $V_{1}$ and $V_{2}$ might be
Higgs-dependent and then they play the role of the Higgs
pre-potentials, but this will be done in the future publication.

\subsection{Equations of motion in the Einstein
frame. }

In Appendix A  we present the equations of motion resulting from
the action (\ref{totaction}) when using the original set of
variables.  The common feature of all the equations in the
original frame is that the measure $\Phi$ degrees of freedom
appear only through dependence upon the scalar field $\zeta$,
Eq.(\ref{zeta}). In particular, all the equations of motion and
the solution for the connection coefficients include terms
proportional to $\partial_{\mu}\zeta$, that makes space-time non
Riemannian and all equations of motion - noncanonical.

It turns out that when working with the new metric ($\phi$
 remains the same)
\begin{equation}
\tilde{g}_{\mu\nu}=e^{\alpha\phi/M_{p}}(\zeta +b_{g})g_{\mu\nu},
\label{ct}
\end{equation}
which we call the Einstein frame,
 the connection  becomes Riemannian. Since
$\tilde{g}_{\mu\nu}$ is invariant under the scale transformations
(\ref{st}), spontaneous breaking of the scale symmetry (by means
of Eq.(\ref{varphi}) which for our model (\ref{totaction}) takes
the form (\ref{app1})) is reduced in the Einstein frame to the
{\it spontaneous breakdown of the shift symmetry}
\begin{equation}
 \phi\rightarrow\phi +const.
 \label{phiconst}
\end{equation}

Notice that the Goldstone theorem generically is not applicable in
this model\cite{G1}. The reason is the following. In fact, the
scale symmetry (\ref{st}) leads to a conserved dilatation current
$j^{\mu}$. However, for example in the spatially flat FRW universe
the spatial components of the current $j^{i}$ behave as
$j^{i}\propto M^4x^i$ as $|x^i|\rightarrow\infty$. Due to this
anomalous behavior at infinity, there is a flux of the current
leaking to infinity, which causes the non conservation of the
dilatation charge. The absence of the latter implies that one of
the conditions necessary for the Goldstone theorem is missing. The
non conservation of the dilatation charge is similar to the well
known effect of instantons in QCD where singular behavior in the
spatial infinity leads to the absence of the Goldstone boson
associated to the $U(1)$ symmetry.

 After the change of
variables  to the Einstein frame (\ref{ct}) and some simple
algebra, Eq.(\ref{app4}) takes the standard GR form
\begin{equation}
G_{\mu\nu}(\tilde{g}_{\alpha\beta})=\frac{\kappa}{2}T_{\mu\nu}^{eff}
 \label{gef}
\end{equation}
where  $G_{\mu\nu}(\tilde{g}_{\alpha\beta})$ is the Einstein
tensor in the Riemannian space-time with the metric
$\tilde{g}_{\mu\nu}$; the energy-momentum tensor
$T_{\mu\nu}^{eff}$ is now
\begin{eqnarray}
T_{\mu\nu}^{eff}&=&\frac{\zeta +b_{\phi}}{\zeta +b_{g}}
\left(\phi_{,\mu}\phi_{,\nu}-\frac{1}{2}
\tilde{g}_{\mu\nu}\tilde{g}^{\alpha\beta}\phi_{,\alpha}\phi_{,\beta}\right)
-\tilde{g}_{\mu\nu}\frac{b_{g}-b_{\phi}}{2(\zeta +b_{g})}
\tilde{g}^{\alpha\beta}\phi_{,\alpha}\phi_{,\beta}
+\tilde{g}_{\mu\nu}V_{eff}(\phi;\zeta,M)
 \label{Tmn}
\end{eqnarray}
where the function $V_{eff}(\phi;\zeta,M)$ is defined as
following:
\begin{equation}
V_{eff}(\phi;\zeta ,M)=
\frac{b_{g}\left[M^{4}e^{-2\alpha\phi/M_{p}}+V_{1}\right]
-V_{2}}{(\zeta +b_{g})^{2}}. \label{Veff1}
\end{equation}
 Putting $M$ in the
arguments of $V_{eff}$ we indicate explicitly that $V_{eff}$
incorporates our choice for the integration constant $M$ that
appears as a result of the spontaneous breakdown of the scale
symmetry. We will see in the next sections that $\zeta$-dependence
of $V_{eff}(\phi;\zeta ,M)$ in the form of {\it square} of $(\zeta
+b_{g})^{-1}$ has a key role in the resolution of the old CC
problem in TMT. The reason that only such $\zeta$-dependence
emerges in $V_{eff}(\phi;\zeta ,M)$, and a $\zeta$-dependence is
absent for example in the numerator of $V_{eff}(\phi;\zeta ,M)$,
is a direct result of our basic assumption that $L_1$ and $L_2$
are independent of the measure fields (see item 2 in Sec.II).

The dilaton $\phi$ field equation (\ref{phi-orig}) in the Einstein
frame reads
\begin{eqnarray}
&&\frac{1}{\sqrt{-\tilde{g}}}\partial_{\mu}\left[\frac{\zeta
+b_{\phi}}{\zeta
+b_{g}}\sqrt{-\tilde{g}}\tilde{g}^{\mu\nu}\partial_{\nu}\phi\right]
\nonumber\\
 &-&\frac{\alpha}{M_{p}(\zeta +b_{g})^{2}} \left[(\zeta
+b_{g})M^{4}e^{-2\alpha\phi/M_{p}}-(\zeta -b_{g})V_{1}
-2V_{2}-\delta b_{g}(\zeta
+b_{g})\frac{1}{2}\tilde{g}^{\alpha\beta}\phi_{,\alpha}\phi_{,\beta}\right]
=0
 \label{phief}
\end{eqnarray}

The scalar field $\zeta$ in Eqs.(\ref{Tmn})-(\ref{phief}) is
determined by means of the constraint (\ref{app3}) which in the
Einstein frame (\ref{ct}) takes the form
\begin{eqnarray}
&&(b_{g}-\zeta)\left[M^{4}e^{-2\alpha\phi/M_{p}}+
V_{1}\right]-2V_{2}-\delta\cdot b_{g}(\zeta +b_{g})X
=0\label{constraint2}
\end{eqnarray}
where
\begin{equation}
X\equiv\frac{1}{2}\tilde{g}^{\alpha\beta}\phi_{,\alpha}\phi_{,\beta}
\qquad \text{and} \qquad \delta =\frac{b_{g}-b_{\phi}}{b_{g}}
\label{delta}
\end{equation}

Applying the constraint (\ref{constraint2}) to Eq.(\ref{phief})
one can reduce the latter to the form
\begin{equation}
\frac{1}{\sqrt{-\tilde{g}}}\partial_{\mu}\left[\frac{\zeta
+b_{\phi}}{\zeta
+b_{g}}\sqrt{-\tilde{g}}\tilde{g}^{\mu\nu}\partial_{\nu}\phi\right]-\frac{2\alpha\zeta}{(\zeta
+b_{g})^{2}M_{p}}M^{4}e^{-2\alpha\phi/M_{p}} =0,
\label{phi-after-con}
\end{equation}
where $\zeta$  is a solution of the constraint
(\ref{constraint2}).

The effective energy-momentum tensor (\ref{Tmn}) can be
represented in a form of that of  a perfect fluid
\begin{equation}
T_{\mu\nu}^{eff}=(\rho +p)u_{\mu}u_{\nu}+p\tilde{g}_{\mu\nu},
\qquad \text{where} \qquad
u_{\mu}=\frac{\phi_{,\mu}}{(2X)^{1/2}}\label{Tmnfluid}
\end{equation}
with the following energy and pressure densities resulting from
Eqs.(\ref{Tmn}) and (\ref{Veff1}) after inserting the solution
$\zeta =\zeta(\phi,X;M)$ of Eq.(\ref{constraint2}):
\begin{equation}
\rho(\phi,X;M) =X+ \frac{(M^{4}e^{-2\alpha\phi/M_{p}}+V_{1})^{2}-
2\delta b_{g}(M^{4}e^{-2\alpha\phi/M_{p}}+V_{1})X -3\delta^{2}
b_{g}^{2}X^2}{4[b_{g}(M^{4}e^{-2\alpha\phi/M_{p}}+V_{1})-V_{2}]},
\label{rho1}
\end{equation}
\begin{equation}
p(\phi,X;M) =X- \frac{\left(M^{4}e^{-2\alpha\phi/M_{p}}+V_{1}+
\delta b_{g}X\right)^2}
{4[b_{g}(M^{4}e^{-2\alpha\phi/M_{p}}+V_{1})-V_{2}]}. \label{p1}
\end{equation}

In a spatially flat FRW universe with the metric
$\tilde{g}_{\mu\nu}=diag(1,-a^2,-a^2,-a^2)$ filled with the
homogeneous scalar sector field $\phi$, the $\phi$  field equation
of motion takes the form
\begin{equation}
Q_{1}\ddot{\phi}+ 3HQ_{2}\dot{\phi}- \frac{\alpha}{M_{p}}Q_{3}
M^{4}e^{-2\alpha\phi/M_{p}}=0 \label{phi1}
\end{equation}
 where $H$ is the Hubble parameter and we have used the following notations
\begin{equation}
\dot{\phi}\equiv \frac{d\phi}{dt}, \label{phidot-vdot}
\end{equation}
\begin{equation}
Q_1=2[b_{g}(M^{4}e^{-2\alpha\phi/M_{p}}+V_{1})-V_{2}]\rho_{,X}
=(b_{g}+b_{\phi})(M^{4}e^{-2\alpha\phi/M_{p}}+V_{1})-
2V_{2}-3\delta^{2}b_{g}^{2}X \label{Q1}
\end{equation}
\begin{equation}
Q_2=2[b_{g}(M^{4}e^{-2\alpha\phi/M_{p}}+V_{1})-V_{2}]p_{,X}=
(b_{g}+b_{\phi})(M^{4}e^{-2\alpha\phi/M_{p}}+V_{1})-
2V_{2}-\delta^{2}b_{g}^{2}X\label{Q2}
\end{equation}
\begin{eqnarray}
-\frac{\alpha}{M_{p}}Q_{3}
M^{4}e^{-2\alpha\phi/M_{p}}&=&2[b_{g}(M^{4}e^{-2\alpha\phi/M_{p}}+V_{1})-V_{2}]\rho_{,\phi}
=\frac{1}{[b_{g}(M^{4}e^{-2\alpha\phi/M_{p}}+V_{1})-V_{2}]}
\nonumber
\\
 &\times&  \left[(M^{4}e^{-2\alpha\phi/M_{p}}+V_{1})
[b_{g}(M^{4}e^{-2\alpha\phi/M_{p}}+V_{1})-2V_{2}] +2\delta
b_{g}V_{2}X+3\delta^{2}b_{g}^{3}X^{2}\right]. \label{Q3}
\end{eqnarray}

It is interesting that the non-linear $X$-dependence appears here
in the framework of the fundamental theory without exotic terms in
the Lagrangians $L_1$ and $L_2$, see Eqs.(\ref{S}) and
(\ref{totaction}).  This effect follows just from the fact that
there are no reasons to choose the parameters $b_{g}$ and
$b_{\phi}$ in the action (\ref{totaction}) to be equal in general;
on the contrary, the choice $b_{g}=b_{\phi}$ would be a fine
tuning. Besides one should stress that the $\phi$ dependence in
$\rho$, $p$ and in equations of motion emerges only in the form
$M^{4}e^{-2\alpha\phi/M_{p}}$ where $M$ is the integration
constant (see Eq.(A1)), i.e. due to the spontaneous breakdown of
the scale symmetry (\ref{st}) (or the shift symmetry
(\ref{phiconst}) in the Einstein frame). Thus the above equations
represent an {\it explicit example of $k$-essence}\cite{k-essence}
{\it resulting from first principles}. The system of equations
(\ref{gef}), (\ref{rho1})-(\ref{phi1}) accompanied with the
functions (\ref{Q1})-(\ref{Q3}) and written in the metric
$\tilde{g}_{\mu\nu}=diag(1,-a^2,-a^2,-a^2)$ can be obtained from
the k-essence type effective action
\begin{equation}
S_{eff}=\int\sqrt{-\tilde{g}}d^{4}x\left[-\frac{1}{\kappa}R(\tilde{g})
+p\left(\phi,X;M\right)\right] \label{k-eff},
\end{equation}
where $p(\phi,X;M)$ is given by Eq.(\ref{p1}). In contrast to the
simplified models studied in literature\cite{k-essence}, it is
impossible here to represent $p\left(\phi,X;M\right)$ in a
factorizable form like $K(\phi)\tilde{p}(X)$.

\section{Cosmological Dynamics in  Fine Tuned $\delta =0$ Models
.}
\subsection{Equations of motion}

 The qualitative
analysis of equations is significantly simplified if $\delta =0$.
This is what we will assume in this section. Although it looks
like a fine tuning of the parameters (i.e. $b_{g}=b_{\phi}$), it
allows us to understand qualitatively the basic features of the
model. In fact, only in the case $\delta =0$ the effective action
(\ref{k-eff}) takes the form of that of the  scalar field without
higher powers of derivatives. Role of $\delta\neq 0$ in a
possibility to produce an  effect of a super-accelerated expansion
of the late time universe will be studied in Sec.V.

\begin{figure}[htb]
%\begin{center}
\includegraphics[width=18.0cm,height=12cm]{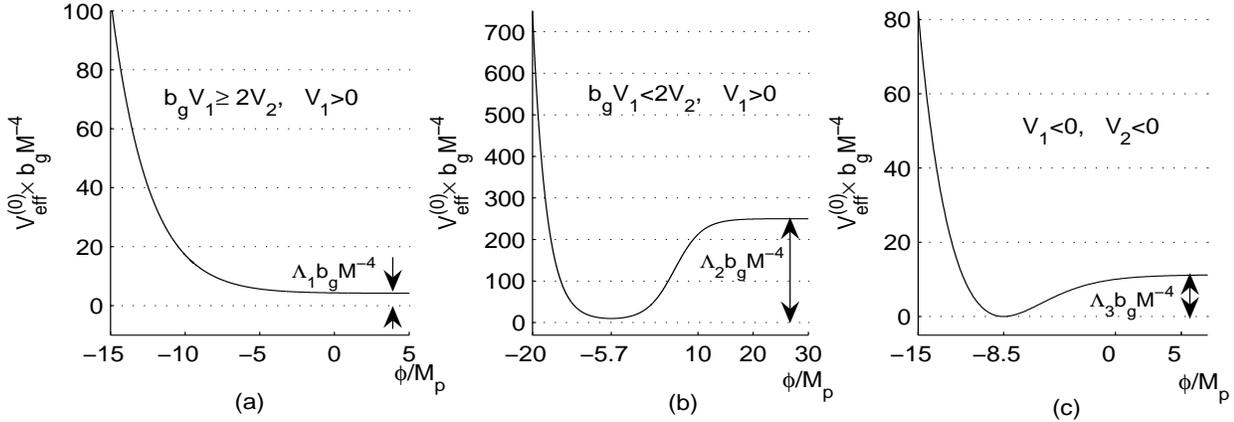}
%\end{center}
\caption{Three possible shapes of the effective potential
$V_{eff}^{(0)}(\phi)$
 in the models with  $b_{g}V_{1}>V_{2}$:
 \quad Fig.(a) \, $b_{g}V_{1}\geq 2V_{2}$ (in the graph
 $V_{1}=10M^{4}$ and $V_{2}=4b_{g}M^{4}$); \quad Fig.(b) \, $b_{g}V_{1}<2V_{2}$
 (in the graph $V_{1}=10M^{4}$ and $V_{2}=9.9b_{g}M^{4}$).
 The value of $V_{eff}^{(0)}$ in the minimum $\phi_{min}=-5.7M_p$
 is larger than zero;
 \quad Fig.(c) \, $V_{1}<0$, $V_{2}<0$
 (in the graph $V_{1}=-30M^{4}$ and $V_{2}=-50b_{g}M^{4}$).
 $V_{eff}^{(0)}(\phi_{min})=0$ in the
 minimum $\phi_{min}=-8.5M_p$ .
In all the cases here as well as in all solutions presented in
this paper we choose $\alpha =0.2$.}\label{fig1}
\end{figure}

So let us study spatially flat FRW cosmological models governed by
the system of equations
\begin{equation}
\frac{\dot{a}^{2}}{a^{2}}=\frac{1}{3M_{p}^{2}}\rho
\label{cosm-phi}
\end{equation}
and (\ref{rho1})-(\ref{phi1}) where one should  set $\delta =0$.

In the fine tuned case under consideration,   the constraint
(\ref{constraint2}) yields
\begin{equation}
\zeta =\zeta(\phi,X;M)|_{\delta =0}\equiv b_{g}-\frac{2V_{2}}
{V_{1}+M^{4}e^{-2\alpha\phi/M_{p}}},
\label{zeta-without-ferm-delta=0}
 \end{equation}
 The energy density and pressure take then the canonical form,
\begin{equation}
\rho|_{\delta =0}=\frac{1}{2}\dot{\phi}^{2}+V_{eff}^{(0)}(\phi);
\qquad p|_{\delta
=0}=\frac{1}{2}\dot{\phi}^{2}-V_{eff}^{(0)}(\phi),
\label{rho-delta=0}
\end{equation}
where the effective potential of the scalar field $\phi$ results
from Eq.(\ref{Veff1})
\begin{equation}
V_{eff}^{(0)}(\phi)\equiv V_{eff}(\phi;\zeta ,M)|_{\delta =0}
=\frac{[V_{1}+M^{4}e^{-2\alpha\phi/M_{p}}]^{2}}
{4[b_{g}\left(V_{1}+M^{4}e^{-2\alpha\phi/M_{p}}\right)-V_{2}]}
\label{Veffvac-delta=0}
\end{equation}
and the $\phi$-equation (\ref{phi1}) is reduced to
\begin{equation}
\ddot{\phi}+3H\dot{\phi}+\frac{dV^{(0)}_{eff}}{d\phi}=0.
\label{eq-phief-without-ferm-delta=0}
\end{equation}

Notice that $V_{eff}^{(0)}(\phi)$ is non-negative for any $\phi$
provided
\begin{equation}
b_{g}V_{1}\geq V_{2} \label{bV1>V2},
\end{equation}
that we will assume in this paper. We assume also that $b_{g}>0$.

In the following three subsections  we consider three different
dilaton-gravity cosmological models determined by different choice
of the parameters $V_{1}$ and $V_{2}$: one model with $V_{1}<0$
and two models with $V_{1}>0$. The appropriate three possible
shapes of the effective potential $V_{eff}^{(0)}(\phi)$ are
presented in Fig.1. A special case with the fine tuned condition
$b_{g}V_{1}=V_{2}$ is discussed in Appendix B where we show that
equality of the couplings to measures $\Phi$ and $\sqrt{-g}$ in
the action (equality $b_{g}V_{1}=V_{2}$ is one of the conditions
for this to happen) gives rise to a symmetric form of the
effective potential.

\subsection{Model
with $V_{1}<0$ and $V_{2}<0$}
 \subsubsection{Resolution of the Old Cosmological Constant Problem in TMT}

The most remarkable feature of the effective potential
(\ref{Veffvac-delta=0}) is that it is proportional to the square
of $V_{1}+ M^{4}e^{-2\alpha\phi/M_{p}}$ which is a straightforward
consequence of our basic assumption that $L_1$ and $L_2$ are
independent of the measure fields (see item 2 in Sec.II,
Eq.(\ref{Veff1}) and the discussion after it). Due to this, as
$V_{1}<0$ and $V_{2}<0$, {\it the effective potential has a
minimum where it equals zero automatically}, without any further
tuning of the parameters $V_{1}$ and $V_{2}$ (see also Fig.1c).
This occurs in the process of evolution of the field $\phi$ at the
value of $\phi =\phi_{0}$ where
\begin{equation}
V_{1}+ M^{4}e^{-2\alpha\phi_{0}/M_{p}}=0 \label{Veff=0}.
\end{equation}
This means that the universe evolves into the state with zero
cosmological constant without any additional tuning of the
parameters  and initial conditions.

\begin{figure}[htb]
\begin{center}
\includegraphics[width=15.0cm,height=8.0cm]{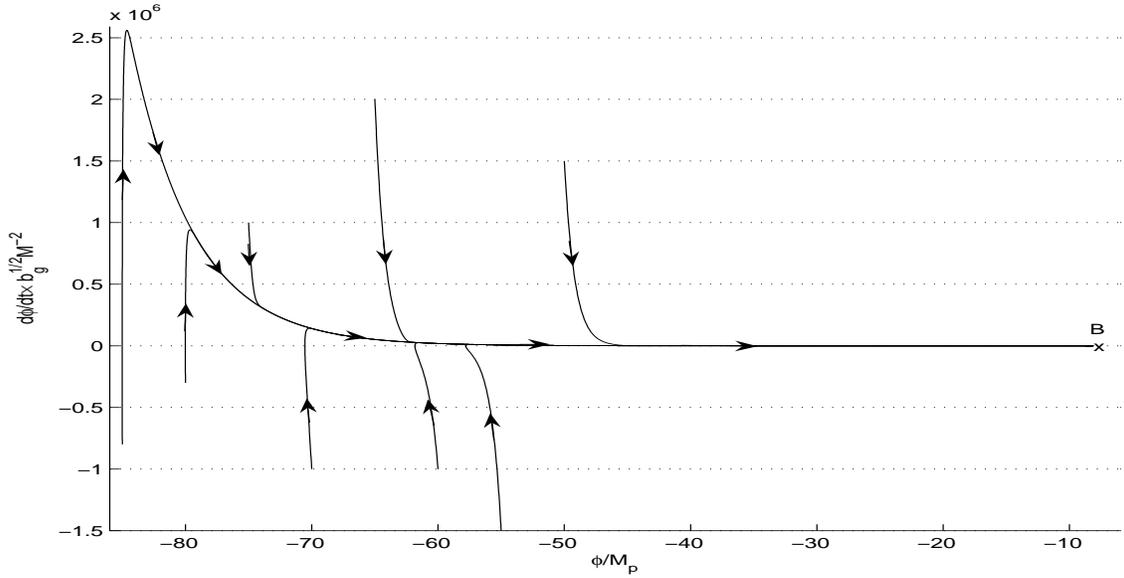}
\end{center}
\caption{Phase portrait (plot of $\frac{d\phi}{dt}$ versus $\phi$)
for the model with $V_{1}<0$ and $V_{2}<0$.   All phase curves
started with $|\phi|\gg M_{p}$ quickly approach the attractor long
before entering the oscillatory regime. The region of the
oscillatory regime is marked by point $B$. The oscillation spiral
is not visible here because of the chosen scales along the
axes.}\label{fig9}
\end{figure}

\begin{figure}[htb]
\begin{center}
\includegraphics[width=16.0cm,height=7.0cm]{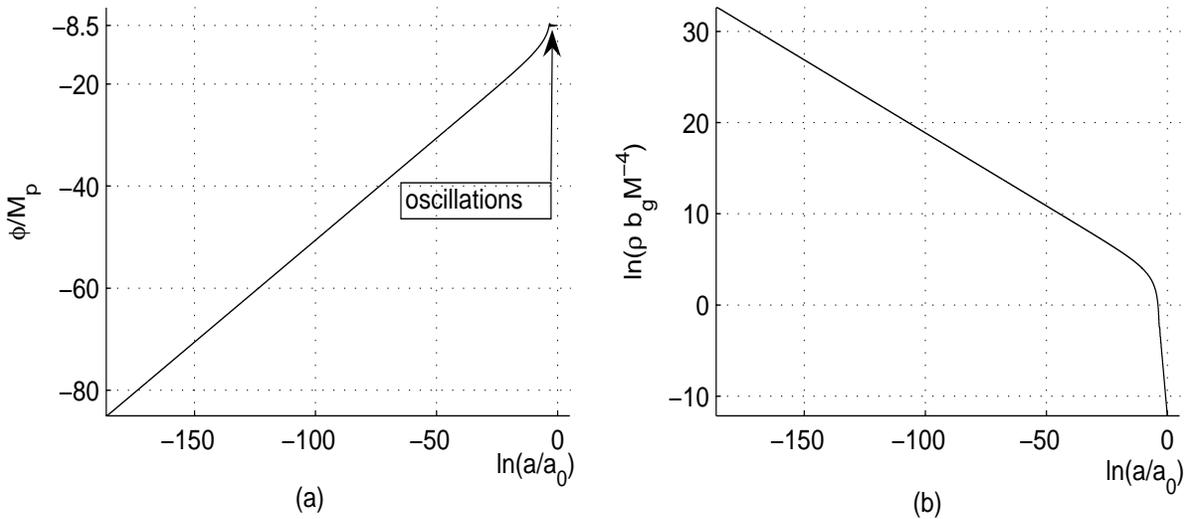}
\end{center}
\caption{(a) Typical dependence of the field $\phi$ (fig. (a)) and
the energy density $\rho$ (fig. (b)) upon $\ln(a/a_{0})$.  Here
and in all the graphs of this paper describing scale factor $a$
dependences, $a(t)$ is normalized such that at the end point of
the described process $a(t_{end})=a_{0}$. In the model with
$V_{1}<0$ the power law inflation ends with damped oscillations of
$\phi$ around $\phi_{0}$ determined by Eq.(\ref{Veff=0}). For the
choice $V_{1}=-30M^{4}$ Eq.(\ref{Veff=0}) gives $\phi_{0}=-8.5M_p$
. (b) The exit from the early inflation is accompanied with
approaching zero of the energy density $\rho$. The graphs
correspond to the evolution which starts from the initial
  values  $\phi_{in} =-85M_{p}$, $\dot{\phi}_{in}=
  -8\cdot 10^5M^2/\sqrt{b_g}$. }\label{fig10}
\end{figure}

\begin{figure}[htb]
\includegraphics[width=8.5cm,height=8.0cm]{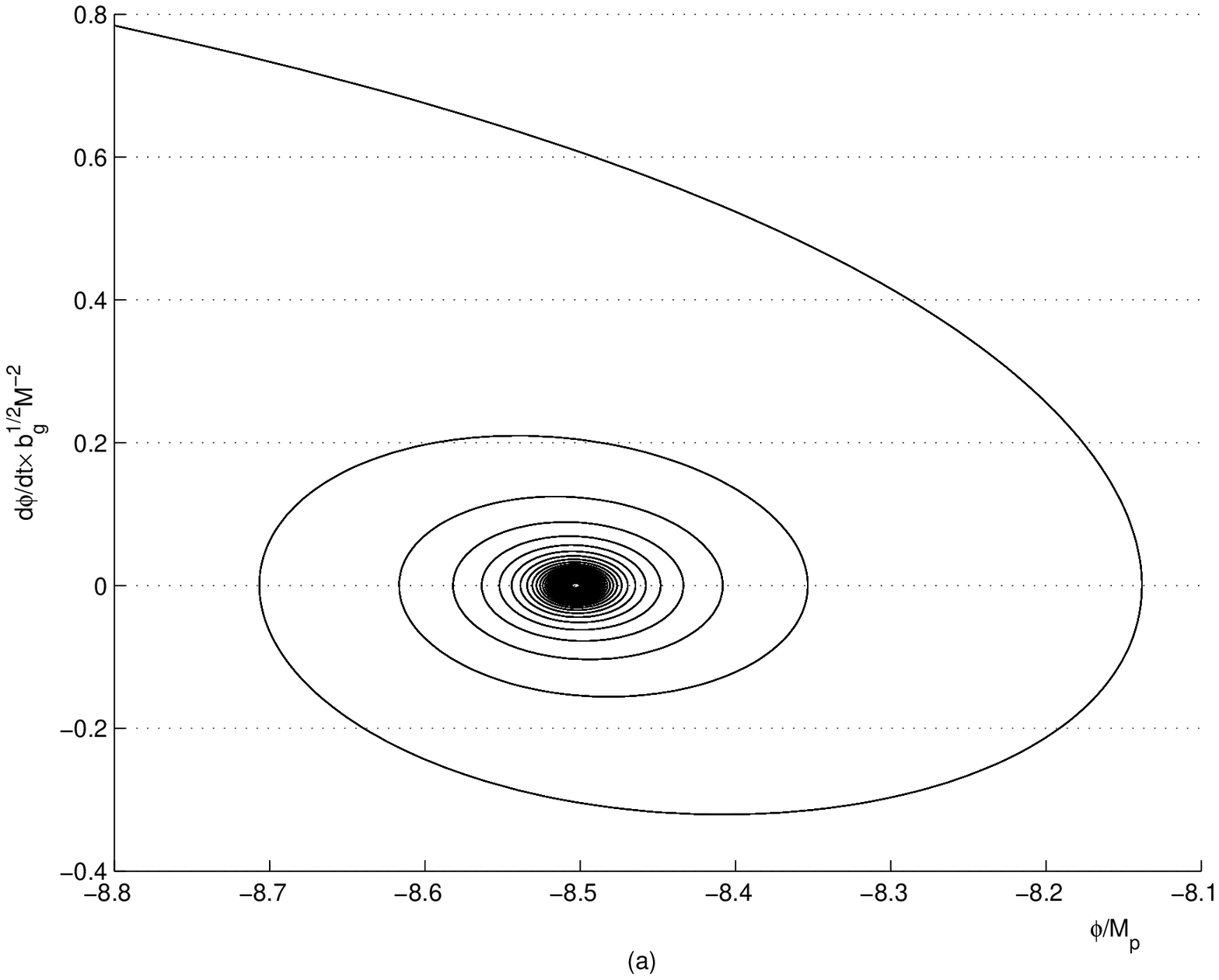}
\includegraphics[width=8.5cm,height=8.0cm]{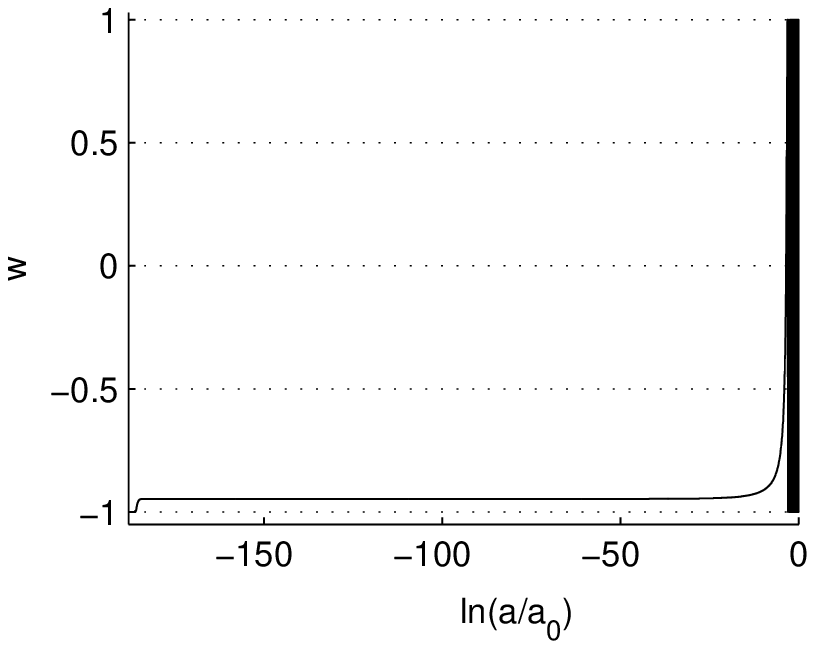}
\caption{Fig.(a) zoom in on the oscillatory regime which marked by
point B in Fig.2. \, (b) Equation-of-state $w=p/\rho$ as function
of the scale factor for the parameters and initial conditions as
in Fig.3. Most of the time the expansion of the universe is a
power law inflation with almost constant $w\approx -0.95$; $w$
oscillates between $-1$ and $1$ at the exit from inflation stage,
i.e. as $\phi\rightarrow\phi_{0}$ and $\rho\rightarrow
0$.}\label{fig11}
\end{figure}

To provide the global scale invariance (\ref{st}), the
prepotentials $V_1$ and $V_2$ enter in the action
(\ref{totaction}) with factor $e^{2\alpha\phi/M_p}$. However, if
quantum effects (considered in the original frame) break the scale
invariance of the action via modification of existing
prepotentials or by means of generation of other prepotentials
with arbitrary $\phi$ dependence (and in particular a "normal"
cosmological constant term $\int \tilde{\Lambda}\sqrt{-g}d^4x$),
this cannot change the result of TMT that the effective  potential
generated in the Einstein frame is proportional to a perfect
square. Note that the assumption of scale invariance is not
necessary for the effect of appearance of  the perfect square in
the effective potential in the Einstein frame and therefore for
the described mechanism of disappearance of the cosmological
constant, see Refs.\cite{GK2}-\cite{G1}.

 If such type of the structure for the scalar field potential in a
conventional (non TMT) model would be chosen "by hand" it would be
a sort of fine tuning. But in our TMT model it is not the starting
point, {\it it is rather a result} obtained in the Einstein frame
of TMT models with spontaneously broken global scale symmetry
including the shift symmetry $\phi\rightarrow \phi +const$.

On the first glance this effect contradicts the Weinberg's no-go
theorem\cite{Weinberg1} which states that there cannot exist a
field theory model where the cosmological constant is zero without
fine tuning. In Sec.VIB we will study in detail the manner our TMT
model avoids this theorem.

\subsubsection{Cosmological Dynamics}

As $M^{4}e^{-2\alpha\phi/M_{p}}\gg Max\left(|V_{1}|,
|V_{2}|/b_{g}\right)$, the effective potential
(\ref{Veffvac-delta=0}) behaves as the exponential potential
$V_{eff}^{(0)}\approx
\frac{1}{4b_{g}}M^{4}e^{-2\alpha\phi/M_{p}}$. So, as $\phi\ll
-M_{p}$  the model describes the well studied power law inflation
of the early universe\cite{power-law} if $\alpha < 1/\sqrt{2}$:
\begin{equation}
a(t)=a_{in}t^{1/2\alpha^{2}}, \qquad \phi =\phi_{in}
+\frac{M_p}{\alpha}\ln\frac{t}{t_{in}}, \label{p-l-solution}
\end{equation}

Further behavior of the solutions in the model of Sec.IVA with
$V_{1}<0$ and $V_{2}<0$, i.e. for the potential of Fig.1c,  is
qualitatively evident enough. With the choice $\alpha =0.2$,
$V_{1}=-30M^{4}$ and $V_{2}=-50b_{g}M^{4}$, the results of
numerical solutions are presented in Figs. 2,3,4. The phase curves
in Fig.2 demonstrate the well known attractor behavior of the
solutions\footnote{Note that the solution (\ref{p-l-solution}) may
be nonapplicable for a relatively short period of time from the
very beginning where $\frac{d\phi}{dt}$ may be negative that we
observe in Fig.2.} describing the power law
inflation\cite{Halliwell}.

Exit from the inflation starts as $\phi$ becomes close to $\phi_0$
determined by Eq.(\ref{Veff=0}). Then the energy density starts
 to tend to zero very fast, Fig.3b. The exit from the inflation
 occurs when all the
phase trajectories corresponding to different initial conditions
with $\phi_{in}\ll -M_{p}$ practically coincide. The process ends
with oscillatory regime, Fig.4a, where $\phi$ performs damped
oscillations around the minimum of the effective potential (see
also Fig.1c).

\subsection{Model with $V_1>0$ and $b_{g}V_{1}\geq
2V_{2}$: Early Power Law Inflation Ending With Small $\Lambda$
Driven Expansion}

\begin{figure}[htb]
\begin{center}
\includegraphics[width=18.0cm,height=6.0cm]{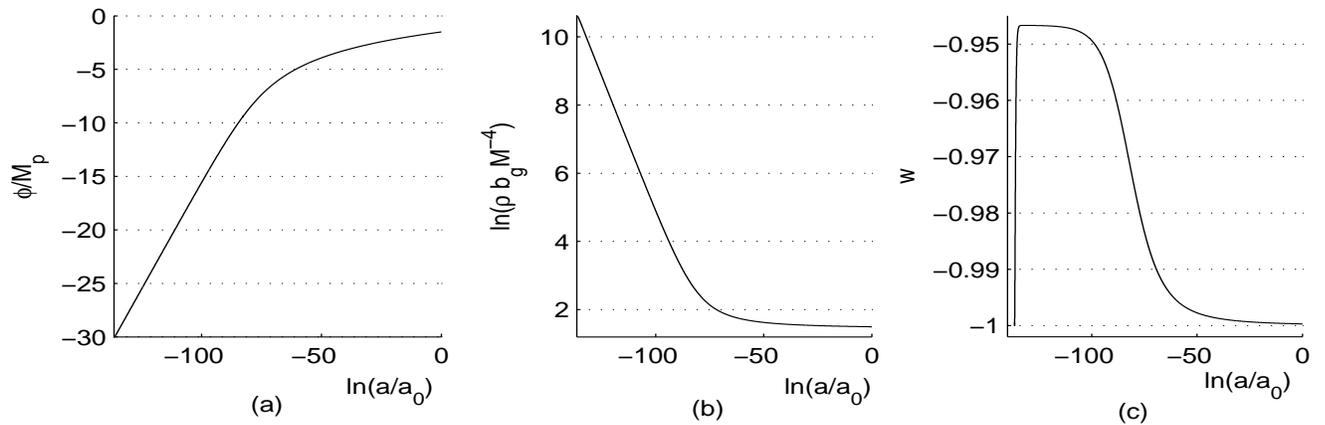}
\end{center}
\caption{ The values of $V_{1}$ and $V_{2}$ are as in Fig.1a. The
graphs correspond to the initial conditions $\phi_{in}=-50M_{p}$,
$\dot{\phi}_{in}=-5M^2b_{g}^{-1/2}$. The early universe evolution
is governed by the almost exponential potential (see Fig.1a)
providing the power low inflation ($w\approx -0.95$ interval in
fig.(c)). After transition to the late time universe the scalar
$\phi$ increases with the rate typical for a quintessence
scenario. Later on  the cosmological constant $\Lambda_{1}$
becomes a dominated component of the dark energy that is displayed
by the infinite region where $w\approx -1$ in
fig.(c).}\label{fig2}
\end{figure}

\begin{figure}[htb]
\begin{center}
\includegraphics[width=14.0cm,height=8.0cm]{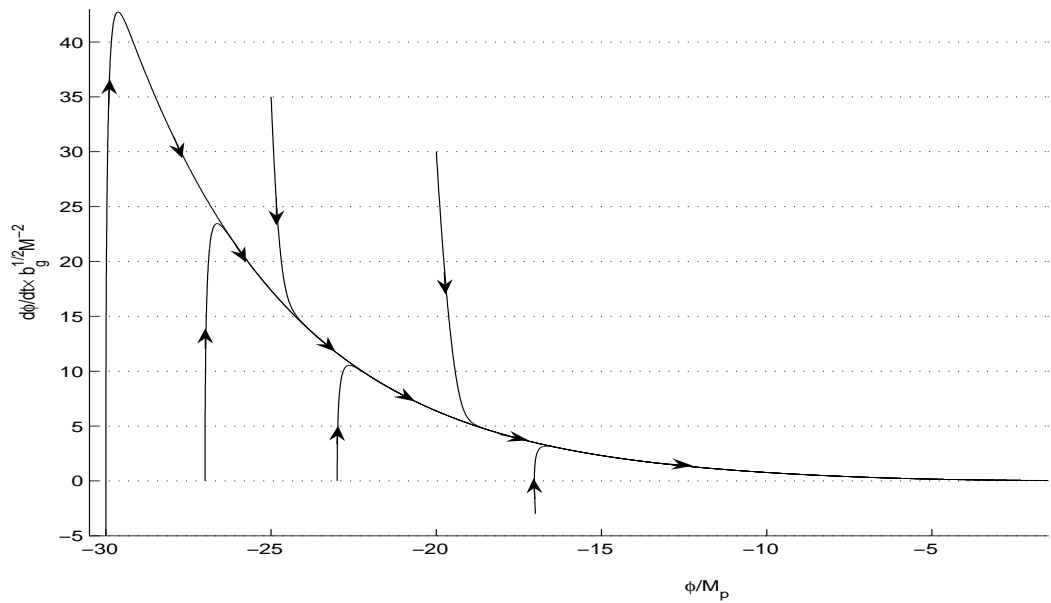}
\end{center}
\caption{Phase portrait (plot of $\frac{d\phi}{dt}$ versus $\phi$)
for the model with $b_{g}V_{1}>2V_{2}$.   All trajectories
 approach the attractor which in its turn
asymptotically (as $\phi\rightarrow\infty$) takes the form of the
straight line $\dot\phi =0$.}\label{fig3}
\end{figure}

 In this model
 the effective potential (\ref{Veffvac-delta=0}) is a
monotonically decreasing function of $\phi$ (see
Fig.{\ref{fig1}}a).  As $\phi\ll -M_{p}$  the model describes the
power law inflation (\ref{p-l-solution}), similar to what we
discussed in the model of previous subsection.

Applying this model to the cosmology of the late time universe and
assuming that the scalar field $\phi\rightarrow\infty$ as
$t\rightarrow\infty$, it is convenient to represent the effective
potential (\ref{Veffvac-delta=0}) in the form
\begin{equation}
V_{eff}^{(0)}(\phi)=\Lambda_1+V_{q-l}(\phi)\qquad
\text{where}\qquad \Lambda_1=\Lambda |_{b_{g}V_1>2V_2},
\label{rho-without-ferm}
\end{equation}
with the definition
\begin{equation}
 \Lambda
=\frac{V_{1}^{2}} {4(b_{g}V_{1}-V_{2})}. \label{lambda}
\end{equation}
Here $\Lambda$ is the positive cosmological constant (see
(\ref{bV1>V2})) and
\begin{equation}
V^{(0)}_{q-l}(\phi)
=\frac{(b_{g}V_{1}-2V_{2})V_{1}M^{4}e^{-2\alpha\phi/M_{p}}+
(b_{g}V_{1}-V_{2})M^{8}e^{-4\alpha\phi/M_{p}}}
{4(b_{g}V_{1}-V_{2})[b_{g}(V_{1}+
M^{4}e^{-2\alpha\phi/M_{p}})-V_{2}]},
\label{V-quint-without-ferm-delta=0}
\end{equation}
that is the evolution of the late time universe  is governed both
by the cosmological constant $\Lambda_1$ and by the
quintessence-like potential $V^{(0)}_{q-l}(\phi)$.

Thus  the effective potential (\ref{Veffvac-delta=0}) provides a
possibility for a cosmological scenario which starts with a power
law inflation and ends with a cosmological constant $\Lambda_1$.
The smallness of $\Lambda_1$ may be achieved without fine tuning
of dimensionfull parameters, that will be discussed in Sec.VI.
Such scenario may be treated as a generalized quintessential
inflation type of scenario. Recall that the $\phi$-dependence of
the effective potential (\ref{Veffvac-delta=0}) appears here only
as the result  of the  spontaneous breakdown of the global scale
symmetry\footnote{The particular case of this model with $b_{g}=0$
and $V_{2}<0$ was studied in Ref.\cite{G1}. The application of the
TMT model with explicitly broken global scale symmetry to the
quintessential inflation scenario was discussed in Ref\cite{K}.}.

Results of numerical solutions for such type of scenario are
presented in Figs.5 and 6 ($V_{1}=10M^{4}$, $V_{2}=4b_{g}M^{4}$)
The early universe evolution is governed by the almost exponential
potential (see Fig.1a) providing the power low inflation
($w\approx -0.95$ interval in fig.(c)) with the attractor behavior
of the solutions, see Ref.\cite{Halliwell}. After transition to
the late time universe the scalar $\phi$ increases with the rate
typical for a quintessence scenario. Later on  the cosmological
constant $\Lambda_{1}$ becomes a dominated component of the dark
energy that is displayed by the infinite region where $w\approx
-1$ in fig.(c). The phase portrait in Fig.3 shows that all the
trajectories started with $|\phi|\gg M_{p}$ quickly approach the
attractor which asymptotically (as $\phi\rightarrow\infty$) takes
the form of the straight line $\dot\phi =0$. Qualitatively similar
results are obtained also when $V_{1}$ is positive but $V_{2}$ is
negative.

\subsection{Model with $V_1>0$ and $V_{2}<b_{g}V_{1}<2V_{2}$}

In this case the effective potential (\ref{Veffvac-delta=0}) has
the minimum (see Fig.1b)
\begin{equation}
V^{(0)}_{eff}(\phi_{min})=\frac{V_{2}}{b_{g}^{2}} \qquad \text{at}
\qquad \phi =\phi_{min}=
-\frac{M_{p}}{2\alpha}\ln\left(\frac{2V_{2}-b_{g}V_{1}}{b_{g}}\right).
\label{minVeff}
\end{equation}

\begin{figure}[htb]
\begin{center}
\includegraphics[width=16.0cm,height=8.0cm]{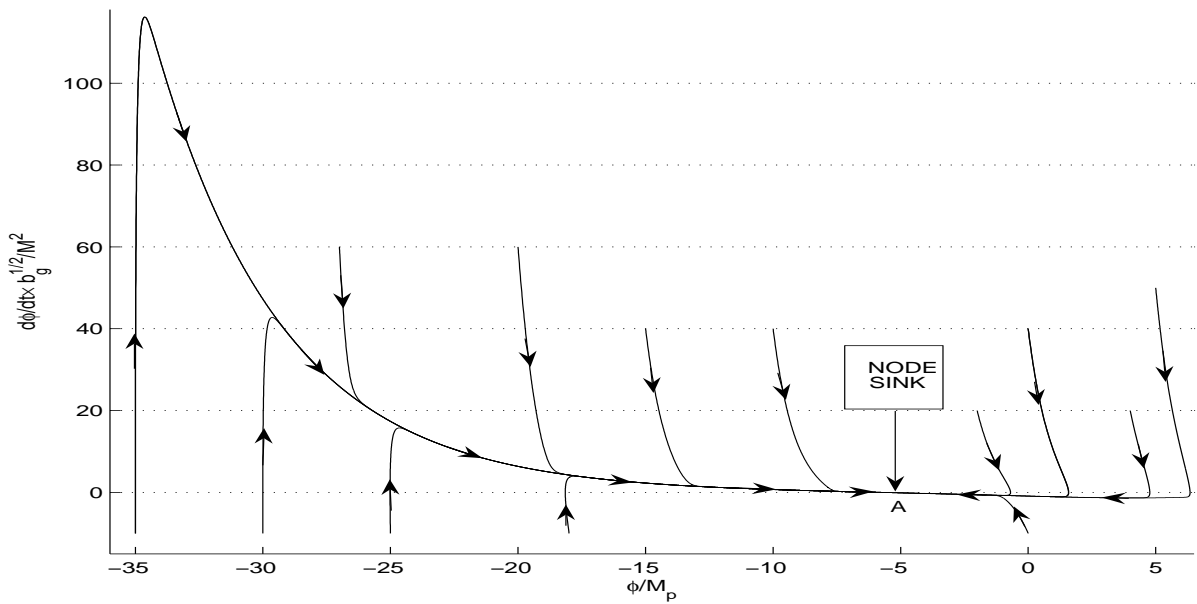}
\end{center}
\caption{Phase portrait (plot of $\frac{d\phi}{dt}$ versus $\phi$)
 for the model with $b_{g}V_{1}<2V_{2}$ and
$V_{1}>0$ (the parameters are chosen here as in Fig.1b ).
Trajectories started anywhere in the phase plane in a finite time
end up at  the same point $A(-5.7,0)$ which is a node sink. There
exist two attractors ending up at $A$, one from the left and other
from the right. All phase curves starting with $|\phi|\gg M_{p}$
quickly approach these attractors.}\label{fig4}
\end{figure}

For the choice of the parameters as in Fig.1b, i.e.
$V_{1}=10M^{4}$ and $V_{2}=9.9b_{g}M^{4}$, the minimum is located
at $\phi_{min}=-5.7M_p$. The character of the phase portrait one
can see in Fig.\ref{fig4}.

\begin{figure}[htb]
\begin{center}
\includegraphics[width=18.0cm,height=6.0cm]{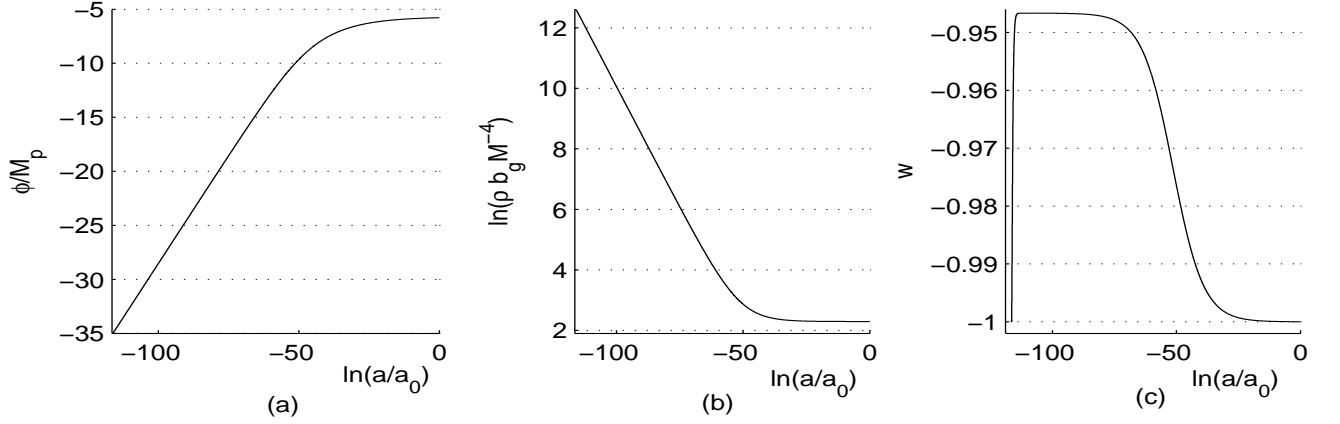}
\end{center}
\caption{Cosmological dynamics in the model with
$b_{g}V_{1}<2V_{2}$ and $V_{1}>0$: typical dependence of $\phi$
(Fig.(a)), the energy density $\rho$ (Fig.(b)) and the
equation-of-state $w$ (Fig.(c)) upon $\ln(a/a_{0})$ where the
scale factor $a(t)$  normalized as in Fig.2. The graphs correspond
to the initial conditions $\phi_{in}=-35M_{p}$,
$\dot{\phi}_{in}=-10b_{g}^{-1/2}M^{2}$. The early universe
evolution is governed by an almost exponential potential (see
Fig.1b) providing the power low inflation ($w\approx -0.95$
interval in Fig.(c)). After arriving the minimum of the potential
at  $\phi_{min}= -5.7M_p$ (see Fig.1b and
 the point $A(-5.7M_p,0)$ of the phase plane in Fig.4) the scalar $\phi$
remains constant. At this stage the dynamics of the universe is
governed by the  constant energy density $\rho
=V^{(0)}_{eff}(\phi_{min})$  (see the appropriate intervals $\rho
=const$ in Fig.(b) and  $w= -1$ in Fig.(c)).}\label{fig5}
\end{figure}

For the early universe as $\phi\ll -M_{p}$, similar to what we
have seen in the models of the previous two subsections, the model
implies the power low inflation. However, the phase portrait
Fig.\ref{fig4} shows that now all solutions end up without
oscillations at the minimum $\phi_{min}=-5.7M_p$ with
$\frac{d\phi}{dt}=0$. In this final state of the scalar field
$\phi$, the evolution of the universe is governed by the
cosmological constant $V^{(0)}_{eff}(\phi_{min})$ determined by
Eq.(\ref{minVeff}). For some details of the cosmological dynamics
see Fig.\ref{fig5}. The desirable smallness of
$V^{(0)}_{eff}(\phi_{min})$ can be provided again without fine
tuning of the dimensionfull parameters that will be discussed in
Sec.VI. The absence of appreciable oscillations in the minimum is
explained by the following two reasons: a) the non-zero friction
at the minimum determined by the cosmological constant
$V^{(0)}_{eff}(\phi_{min})$; b) the shape of the potential near to
minimum is too flat.

The described properties of the model are evident enough.
Nevertheless we have presented them here because this model is a
particular (fine tuned) case of an appropriate model with
$\delta\neq 0$ studied in the next subsection where we will
demonstrate a possibility of states with $w<-1$ without any exotic
contributions, like a phantom term, in the original action.

\section{Dynamics in the Case $\delta\neq 0$: A Superacceleration
Phase of the Universe.}

We return now to the general case of our model (see Sec.III) with
no fine tuning of the parameters $b_g$ and $b_{\phi}$, i.e. the
parameter $\delta$, defined by Eq.(\ref{delta}), is non zero. We
have checked that  the models of Secs.IVB and IVC do not suffer
qualitative changes at all and quantitative modifications are not
strong, at least in the range of $-1<\delta<1$. However the model
of Sec.IVD changes drastically as $\delta\neq 0$: the model
enables k-essence type solutions for the late time universe with
equation-of-state $w<-1$.

\begin{figure}[htb]
\begin{center}
\includegraphics[width=17.0cm,height=12.0cm]{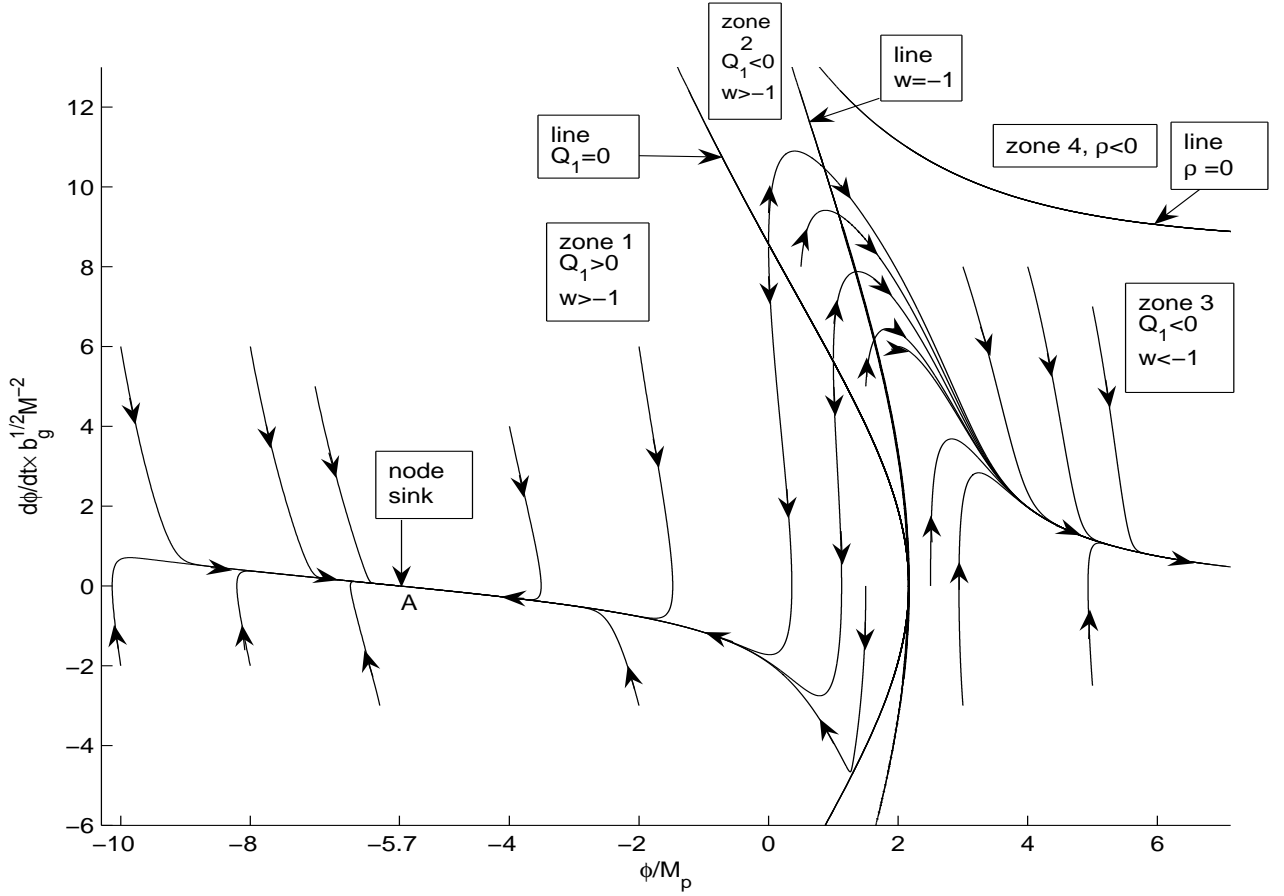}
\end{center}
\caption{The phase portrait for the model with $\alpha =0.2$,
$\delta =0.1$, $V_{1}=10M^{4}$ and $V_{2}=9.9b_{g}M^{4}$. The
region with $\rho >0$ is divided into two dynamically disconnected
regions by the line $Q_{1}(\phi,\dot{\phi})=0$. To the left of
this line $Q_{1}>0$ (the appropriate zone we call zone 1) and to
the right - $Q_{1}<0$. The phase portrait in zone 1 corresponds to
processes similar to those of Sec.IVD. The $\rho >0$ region to the
right of the line $Q_{1}(\phi,\dot{\phi})=0$ is divided into two
zones (zone 2 and zone 3) by the line $Q_2=0$ (the latter
coincides with the line where $w=-1$). In zone 2 \, $w>-1$ but
$c_s^2<0$. In zone 3 \,$w<-1$ and $c_s^2>0$. Phase curves can
start in zone 2 in points very close to the line $Q_1=0$. After
they cross the line $w=-1$, i.e. in zone 3, they exhibit processes
with super-accelerating expansion of the universe. The phase
curves in zone 3 demonstrate dynamical attractor behavior. As
$\phi\rightarrow\infty$ the phase curves approach the straight
line $\dot{\phi}=0$.}\label{fig6}
\end{figure}

\begin{figure}[htb]
\begin{center}
\includegraphics[width=16.0cm,height=5.0cm]{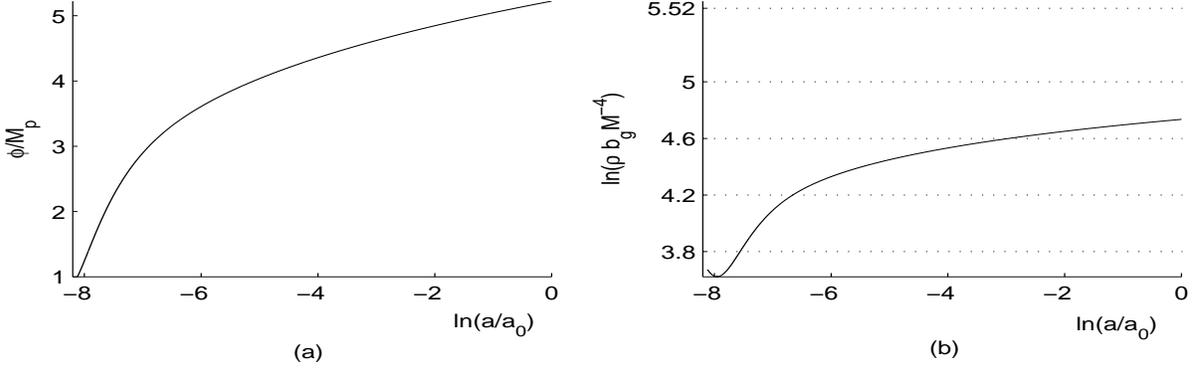}
\end{center}
\caption{ Typical scalar factor dependence of $\phi$  (Fig.(a))
and
 of the energy density $\rho$, defined by Eq.(\ref{rho1}),
(Fig(b)) in the regime corresponding to the phase curves started
in zone 2. Both graphs correspond to the initial conditions
$\phi_{in}=M_{p}$, $\dot\phi_{in} =5.7M^4/\sqrt{b_g}$; $\rho$
increases approaching asymptotically $\Lambda_{2}=
\frac{M^{4}}{b_{g}}e^{5.52}$, where $\Lambda_{2}$ is the value of
$\Lambda$ determined by Eq.(\ref{lambda}) as $V_{1}=10M^{4}$ and
$V_{2}=9.9b_{g}M^{4}$; see also Fig.1b.}\label{fig7}
\end{figure}
\begin{figure}[htb]
\begin{center}
\includegraphics[width=16.0cm,height=5.0cm]{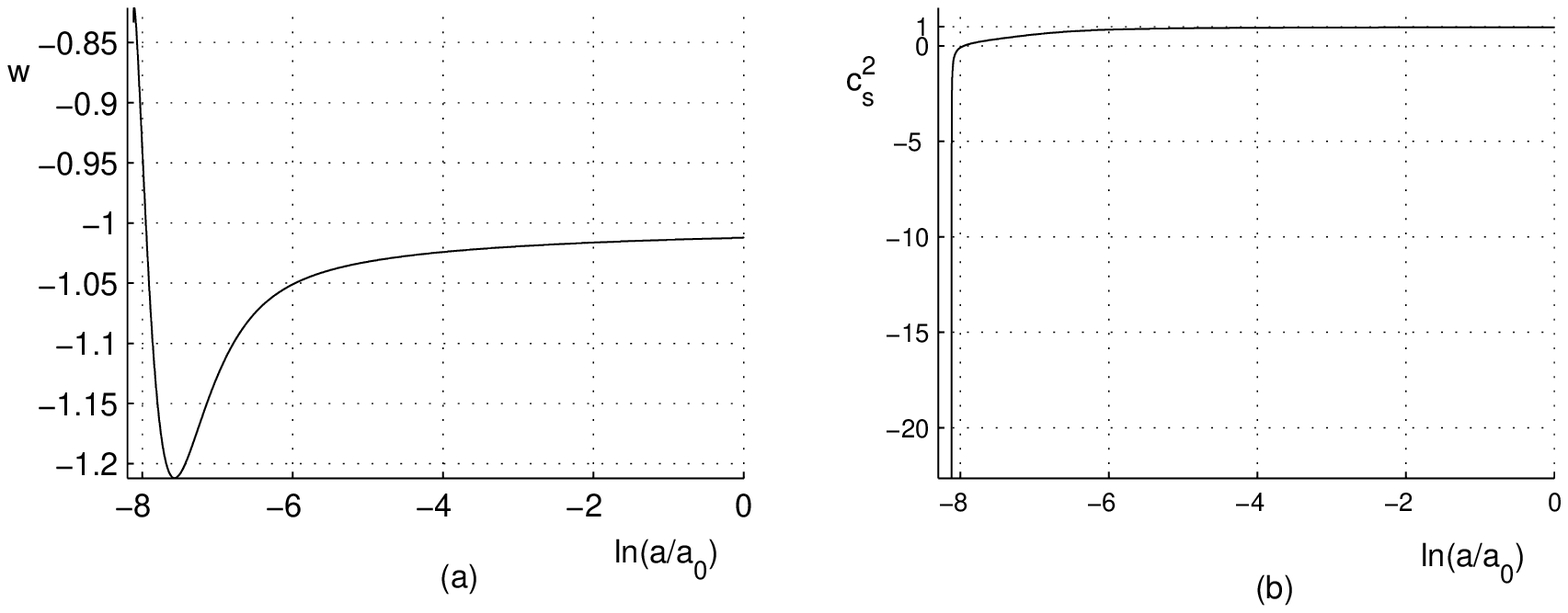}
\end{center}
\caption{The scale factor dependence of the equation-of-state $w$
(Fig.(a)) and effective speed of sound for perturbations $c_s^2$
(Fig.(b)) for the initial conditions $\phi_{in}=M_{p}$,
$\dot\phi_{in} =5.7M^4/\sqrt{b_g}$.} \label{fig8}
\end{figure}

So, we consider now the dynamics of the FRW cosmology described by
Eqs.(\ref{cosm-phi})-(\ref{Q3}). Before choosing the appropriate
parameters for numerical studies, let us start from the analysis
of Eq.(\ref{phi1}). The interesting feature of this equation is
that for certain range of the parameters, each of the factors
$Q_{i}(\phi,X)$ \quad ($i=1,2,3$) can get to zero. Equation
$Q_{i}(\phi,X)=0$ determines a line in the phase plane
$(\phi,\dot\phi)$. In terms of a mechanical interpretation of
Eq.(\ref{phi1}), the change of the sign of $Q_{1}$ can be treated
as the change of the mass of "the particle". Therefore one can
think of situation where "the particle" climbs up in the potential
with acceleration. It turns out that when the scalar field is
behaving in this way, the flat FRW universe may undergo a
super-acceleration.

For $Q_{1}\neq 0$, Eqs. (\ref{phi1}), (\ref{cosm-phi}) result in
the well known equation\cite{Vikman}
\begin{equation}
\ddot{\phi}+
\frac{\sqrt{3\rho}}{M_p}c_s^2\dot{\phi}+\frac{\rho_{,\phi}}{\rho_{,X}}
=0, \label{phantom-phi}
\end{equation}
where $c_s$ is the effective sound speed of
perturbations\cite{k-essence}
\begin{equation}
c_s^2=\frac{p_{,X}}{\rho_{,X}}=\frac{Q_2}{Q_1}, \label{c-s-2}
\end{equation}
\begin{equation}
\frac{\rho_{,\phi}}{\rho_{,X}}=- \frac{\alpha}{M_{p}}\cdot
\frac{Q_3}{Q_1}\cdot M^{4}e^{-2\alpha\phi/M_{p}}; \label{Q-rho-p}
\end{equation}
$\rho$ and $p$ are defined by Eqs.(\ref{rho1}), (\ref{p1}) and
$Q_i$ $(i=1,2,3)$ - by Eqs.(\ref{Q1})-(\ref{Q3}).

With simple algebra one can see that the following "sign rule" is
fulfilled for the equation-of-state $w=p/\rho$:
\begin{equation}
sign(w+1)=sign (Q_2) \label{sgnw+1}
\end{equation}
and the line $Q_2(\phi,X)=0$ divides the phase plane
$\phi,\dot{\phi}$ into two regions: one with $w>-1$ and other with
$w<-1$.

There are a lot of  sets of parameters providing the $w<-1$ phase
in the late universe. For example we are demonstrating here this
effect with the following set of the parameters of the original
action(\ref{totaction}): $\alpha =0.2$, $V_{1}=10M^{4}$ and
$V_{2}=9.9b_{g}M^{4}$ used in Sec.IVD but now we choose $\delta
=0.1$. The results of the numerical solution are presented in
Figs.9-11.

The phase plane, Fig.9, is divided into two regions by the line
$\rho =0$. The region $\rho >0$ is divided into two dynamically
disconnected regions by the line $Q_{1}(\phi,X)=0$.

To the left of the line $Q_{1}(\phi,X)=0$ - zone 1 where
$Q_{1}>0$. Comparing carefully the phase portrait in the zone 1
with that in Fig.\ref{fig4} of the previous subsection, one can
see an effect of $\delta\neq 0$ on the shape of phase
trajectories. However the general structure of these two phase
portraits is very similar. In particular, they have the same node
sink $A(-5.7M_{p},0)$. At this point "the force" equals zero since
$Q_{3}|_A=0$. The value $\phi =-5.7M_{p}$ coincides with the
position of the minimum of $V_{eff}^{(0)}(\phi)$ because in the
limit $\dot{\phi}\rightarrow 0$ the role of the terms proportional
to $\delta$ is negligible. Among trajectories converging to node
$A$ there are also trajectories corresponding to a power low
inflation of the early universe, which is just a generalization to
the case $\delta\neq 0$ of the similar result discussed in the
previous subsection.

In the region to the right of the line $Q_{1}(\phi,X)=0$, all
phase curves approach the attractor which in its turn
asymptotically (as $\phi\rightarrow\infty$) takes the form of the
straight line $\dot\phi =0$. This region is divided into two zones
by the line $Q_2(\phi,X)=0$. In all points of this line $w=-1$. In
zone 2, i.e. between the lines $Q_1=0$ and $w=-1$, the
equation-of-state $w>-1$ and the sound speed $c_s^2<0$. In zone 3,
i.e. between the line $w=-1$ and the line $\rho =0$, the
equation-of-state $w<-1$ and the sound speed $c_s^2>0$.

For a particular choice of the initial data $\phi_{in}=M_{p}$,
$\dot\phi_{in} =9M^4/\sqrt{b_g}$, the features of the solution of
the equations of motion are presented in Figs.10 and 11. The main
features of the solution as we observe from the figures are the
following: 1) $\phi$ slowly increases in time; 2) the energy
density $\rho$ slowly increases approaching the constant $\Lambda
=\Lambda_{2}$  defined by the same formula as in
Eq.(\ref{lambda}), see also Fig.1b; for the chosen parameters
$\Lambda_{2}\approx\frac{M^{4}}{b_{g}}e^{5.52}$; \quad
 3) $w\equiv p/\rho$ is
less than $-1$ and asymptotically approaches $-1$ from below.

Using the classification of Ref.\cite{Vikman} of conditions for
 the dark energy to evolve from the state with $w>-1$ to the
phantom state, we see that transition of the phase curves from
zone 2 (where $w>-1$) to the phantom zone 3 (where $w<-1$) occurs
under the conditions  $p_{,X}=0$, $\rho_{,X}\neq 0$, $X\neq 0$.
Qualitatively the same behavior one observes for all initial
conditions $(\phi_{in},\dot{\phi}_{in})$ disposed in the zone 2.
The question of constructing a realistic scenario where the dark
energy can evolve from the power low inflation state disposed in
zone 1 to the phantom zone 3 is beyond of the goal of this paper.

\section{Resolutions of
the  Cosmological Constant Problems and Connection Between TMT and
Conventional Field Theories with Integration Measure $\sqrt{-g}$}

\subsection{The new cosmological constant problem}
The smallness of the observable cosmological constant $\Lambda$ is
known as the new cosmological constant problem\cite{Weinberg2}. In
TMT, there are two ways to provide the observable order of
magnitude of $\Lambda\sim (10^{-3}eV)^{4}$ by an appropriate
choice of the parameters of the theory (see Eqs.(\ref{lambda}) and
(\ref{bV1>V2})) but {\it without fine tuning of the dimensionfull
parameters}.

\subsubsection{Seesaw mechanism}

 If $V_{2}<0$ then there is no need for $V_{1}$ and $V_{2}$ to be
small: it is enough that $b_{g}V_{1}<|V_{2}|$ and $V_{1}/
|V_{2}|\ll 1 $.  This possibility is a kind of  {\it seesaw}
 mechanism\cite{G1},\cite{seesaw}). For instance, if $V_{1}$ is
determined by the
 energy scale of electroweak symmetry breaking $V_{1}\sim
(10^{3}GeV)^{4}$ and $V_{2}$ is determined by the Planck scale
$V_{2} \sim (10^{18}GeV)^{4}$ then $\Lambda_{1}\sim
(10^{-3}eV)^{4}$. The range of the possible scale of the
dimensionless parameter $b_{g}$ remains very broad.

\subsubsection{The TMT correspondence principle and the smallness of
$\Lambda$}

 Let us start from the notion that if $V_2>0$ or alternatively
 $V_2<0$ and $b_gV_{1}>|V_{2}|$
then $\Lambda\sim \frac{V_1}{b_g}$. Hence the second possibility
to ensure the needed smallness of $\Lambda$ is to choose the {\it
dimensionless} parameter $b_{g}>0$ to be a huge number.   In this
case the order of magnitudes of $V_{1}$ and $V_{2}$ could be
either as in the above case of the seesaw mechanism or to be not
too much different from each other (or even of the same order).
For example, if $V_{1}\sim (10^{3}GeV)^{4}$ then for getting
$\Lambda_1\sim (10^{-3}eV)^{4}$ one should assume that $b_{g}\sim
10^{60}$. Below we will use this value just for illustrative
purposes.

Note that $b_{g}$ is the ratio of the coupling constants of the
scalar curvature to the measures $\sqrt{-g}$ and $\Phi$
respectively in the fundamental action (\ref{totaction}). The
Lagrangians $L_1$ and $L_2$ have the same structure: both of them
contain the scalar curvature, kinetic and pre-potential terms. It
is natural to assume that the ratio of  couplings of all the
appropriate terms in $L_1$ and $L_2$ to the measures $\sqrt{-g}$
and $\Phi$ have the same or close orders of magnitude. This is why
in Sec.III we have made an assumption that the dimensionless
parameters $b_g$ and $b_{\phi}$ have close orders of magnitude.
For the same reason we will also assume that $V_2/V_1\sim b_g\sim
10^{60}$. If this is the case\footnote{Note that if $V_{2}<0$ then
the choice $|V_2|/V_1\sim b_g$ means that in this case the second
way of resolution of the new CC problem is a particular case of
the seesaw mechanism. However the second way is applicable also if
$V_2>0.$} then the huge value of $b_g$ can be treated as an
indication that TMT implies a certain sort of {\it the
correspondence principle} between TMT and conventional field
theories (i.e theories with only the measure of integration
$\sqrt{-g}$  in the action). In fact, using the notations of the
general form of the TMT action (\ref{S}) in the case of the action
(\ref{totaction}), one can conclude that the relation between the
"usual" (i.e. entering in the action with the usual measure
$\sqrt{-g}$) Lagrangian density $L_2$ and the new one $L_1$
(entering in the action with the new measure $\Phi$) is roughly
speaking $L_2\sim 10^{60}L_1$. In the case $\int L_1\Phi d^4x$
becomes negligible, the remaining term of the action $\int
L_2\sqrt{-g} d^4x$ would describe GR instead of TMT. It seems to
be very interesting that such {\it a correspondence principle for
the TMT action (\ref{S}) may have a certain relation to the
extreme smallness of the cosmological constant}.

Appearance of a big dimensionless constant in particle field
theory is usually associated with hierarchy of masses and/or
interactions describing by {\it different} terms in the
Lagrangian. The way big numbers can appear in the TMT action is
absolutely different. It is easier to see this difference in the
case of a fine tuned model, where $b_g=b_{\phi}$ and $b_gV_1=V_2$,
see Appendix B. In such a case the Lagrangians $L_1$ and $L_2$ not
only have the same type of terms but they are just proportional:
$L_2=b_gL_1$. Therefore the nature of the huge value of $b_g$
differs here very much from the conventional hierarchy
issue\footnote{ Note that instead of using factors $b_g$ and
$b_{\phi}$ one can define the fundamental TMT action such that the
appropriate factors in $L_2$ equal unity. Then in front of the
 appropriate terms in $L_1$ one should add factors like $b_g^{-1}$ and
$b_{\phi}^{-1}$. With such a definition of the Lagrangians $L_1$
and $L_2$, instead of a huge factor in $L_2$, a very small factor
will appear in $L_1$.}.

If such the ratio between  $L_1$ and $L_2$ is actually realized,
then taking into account the fact that $L_1$ and $L_2$ describe
the same matter and gravity degrees of freedom in a very similar
manner, the question arises why $L_1$ is not dynamically
negligible in comparison with $L_2$. To answer this question we
have to turn to the fundamental action (\ref{totaction}) that it
is convenient to rewrite in the following form
\begin{eqnarray}
&S=&\int \sqrt{-g}d^{4}x e^{\alpha\phi /M_{p}}
\left[-\frac{b_{g}}{\kappa}R(\Gamma
,g)\left(\frac{\zeta}{b_{g}}+1\right)+\left(\frac{\zeta}{b_{g}}
+\frac{b_{\phi}}{b_{g}}\right)\frac{b_{g}}{2}g^{\mu\nu}\phi_{,\mu}\phi_{,\nu}-e^{\alpha\phi
/M_{p}}\left(\zeta\frac{V_{1}}{V_2} +1\right)V_2\right]
 \label{totaction1}
\end{eqnarray}
where one can see that the ratio $\zeta/b_g$ has an important
dynamical role. Analyzing  the constraint (\ref{constraint2}) and
cosmological dynamics studied in Secs.IV and V it is easy to see
that the order of magnitude of the scalar field
$\zeta\equiv\Phi/\sqrt{-g}$ is generically close to that of $b_g$.
In other words, when analyzing the action on the mass shell, it
turns out that the ratio of the measures $\Phi/\sqrt{-g}$
generically compensates the smallness of $L_1/L_2$. So, similar
terms ($R$-terms, kinetic terms and pre-potential terms) appearing
in the action (\ref{totaction}) with measures $\Phi$ and
$\sqrt{-g}$ respectively, are both dynamically important in
general.

In the light of this understanding of the general picture it is
interesting to check the TMT dynamics in situations where
$\zeta/b_g$ becomes very small or very large. Let us start from
the fine tuned model where $b_gV_1=2V_2$ and $b_g=b_{\phi}$
(recall that $V_1>0$ and $b_g>0$ in all our models). In this case
it follows from the constraint (\ref{zeta-without-ferm-delta=0})
that
\begin{equation}
\frac{\zeta}{ b_{g}}=\frac{M^{4}e^{-2\alpha\phi/M_{p}}}
{V_{1}+M^{4}e^{-2\alpha\phi/M_{p}}}, \label{zeta-special}
 \end{equation}
Then the effective potential (\ref{Veffvac-delta=0}) (see also
Eqs.(\ref{rho-without-ferm})-(\ref{V-quint-without-ferm-delta=0}))
reads
\begin{equation}
V_{eff}^{(0)}(\phi)=\frac{V_2}{b_g^2}+\frac{M^{8}e^{-4\alpha\phi/M_{p}}}
{2b_g(V_1+ 2M^{4}e^{-2\alpha\phi/M_{p}})}. \label{Veff-special}
\end{equation}
In such a fine tuned model, $\zeta/b_g$ approaches zero
asymptotically as $\phi\rightarrow\infty$ where the effective
potential becomes flat. However when looking into the TMT action
written in the form (\ref{totaction1}) we see that the asymptotic
disappearance of $\zeta/b_g$ means that we deal with an asymptotic
transition from TMT to a conventional field theory model with only
measure of integration $\sqrt{-g}$ and only one Lagrangian
density. In the limit $\zeta/b_g\rightarrow 0$, the conformal
transformation to the Einstein frame (\ref{ct}) takes the form
$\tilde{g}_{\mu\nu}=b_{g}e^{\alpha\phi/M_{p}}g_{\mu\nu}$.
Therefore in the Einstein frame, the limit of the action
(\ref{totaction1}) as $\zeta/b_g\rightarrow 0$ is reduced to the
following  model:
\begin{equation}
S|_{\zeta=0,\, Eistein\, frame}=\int \sqrt{-\tilde{g}}d^{4}x\left[
-\frac{1}{\kappa}R(\tilde{g})+\frac{1}{2}\tilde{g}^{\mu\nu}\phi_{,\mu}\phi_{,\nu}-\frac{1}{b_g^2}V_2
\right].
 \label{totaction-zeta0}
\end{equation}
It is easy to see that for example in the  FRW universe, the
asymptotic (as the scale factor $a(t)\rightarrow\infty$) behavior
of the universe in the model (\ref{totaction-zeta0}) coincides
with the appropriate asymptotic result of TMT model under
consideration: both of them asymptotically describe the universe
governed by the cosmological constant $V_2/b_g^2=V_1/2b_g$.

Similar conclusion is obtained in a model where $V_2<b_gV_1<2V_2$,
i.e. with no fine tuning of the prepotentials, which have been
studied in Sec.IVD. The only difference is that now
$\zeta/b_g\rightarrow 0$ and $V_{eff}\rightarrow V_2/b_g^2\sim
V_1/b_g$ as $\phi\rightarrow\phi_{min}$, see Eq.(\ref{minVeff});
in a small neighborhood of $\phi_{min}$ the TMT action presented
in the Einstein frame looks like (\ref{totaction-zeta0}).

Note however that in the context of the k-essence model studied in
Sec.V where $\delta\neq 0$ and  $V_2<b_gV_1<2V_2$, the asymptotic
value of $\zeta$ in the late time universe (i.e. as
$\phi\rightarrow\infty$ and $X\rightarrow 0$) is $|\zeta|\sim b_g$
but nevertheless the energy density tends to $\Lambda =\Lambda_2
=\frac{V_{1}^{2}} {4(b_{g}V_{1}-V_{2})}\sim V_1/b_g$ as well (see
Fig.1b).

 Similar situation takes place in the model where $b_gV_1>2V_2$, studied
in Sec.IVC. The asymptotic value of $\zeta$ in the late time
universe is again $\zeta\sim b_g$  and the energy density tends to
$\Lambda =\Lambda_1 =\frac{V_{1}^{2}} {4(b_{g}V_{1}-V_{2})}\sim
V_1/b_g$ as well (see Fig.1a).

Thus in all the models, the huge value of $b_g$ can ensure the
needed smallness of the dark energy density in the late time
universe but it is not always realized due to the limit
$\zeta/b_g\rightarrow 0$.

\subsection{The Old Cosmological Constant Problem
Is Solved in the Dynamical Regime where the  Fundamental TMT
Action Tends to a Limit Opposite to Conventional Field Theory
(with only measure $\sqrt{-g}$).}

As we have seen in Sec.IVB, in the model with $V_{1}<0$ and
$V_{2}<0$, the old cosmological constant problem is resolved
without fine tuning: the effective potential
(\ref{Veffvac-delta=0}) is proportional to the square of
$V_{1}+M^{4}e^{-2\alpha\phi/M_{p}}$, and  $\phi =\phi_{0}$ where
$V_{1}+ M^{4}e^{-2\alpha\phi_{0}/M_{p}}=0$, is the minimum of the
effective potential without any further tuning of the parameters
and initial conditions. Now we want to analyze some of the
essential differences we have in TMT as compared with the
conditions of the Weinberg's no-go theorem and show what are the
reasons providing solution of the old CC problem in TMT. This has
to be done when TMT is considered in the original frame since in
the Einstein frame we observe only the results in the effective
picture after some of the symmetries are broken.

\begin{itemize}

\item
The basic assumption of the Weinberg's  theorem is that in the
vacuum all the fields (metric tensor $g_{\mu\nu}$ and matter
fields $\psi_n$) are constant. As it was pointed out by S.Weinberg
in the review\cite{Weinberg1}, the Euler-Lagrange equations  for
such constant fields (with the action $\int{\cal L}\left(
g_{\mu\nu},\psi_n \right)d^4x$) have the form
\begin{equation}
\frac{\partial{\cal L}}{\partial g_{\mu\nu}}=0, \label{W1}
\end{equation}
\begin{equation}
\frac{\partial{\cal L}}{\partial\psi_n}=0
 \label{W2}
\end{equation}
and these equations constitute the basis for further Weinberg's
arguments. In particular, if $GL(4)$ symmetry
\begin{equation}
g_{\mu\nu}\rightarrow
A^{\alpha}_{\mu}A^{\beta}_{\nu}g_{\alpha\beta}, \qquad
\psi_i\rightarrow D_{ij}(A)\psi_j \label{W3}
\end{equation}
survives as a vestige of general covariance when all the fields
are constrained to be constant, the Lagrangian ${\cal L}$
transforms as a density:
\begin{equation}
{\cal L}\rightarrow detA\cdot{\cal L}. \label{W4}
\end{equation}
 Weinberg concludes that when
Eq.(\ref{W1}) is satisfied then the unique form of ${\cal L}$ is
\begin{equation}
{\cal L}=c\sqrt{-g}, \label{W5}
\end{equation}
where $c$ is independent of $g_{\mu\nu}$. As a matter of fact this
means that for example in the case of a scalar matter field $\phi$
model considered by Weinberg in Sec.VI of the review
\cite{Weinberg1}, $c$ is determined by the value of the scalar
field $\phi$ potential as $\phi$ is a constant determined by
Eq.(\ref{W2}).

However, if ${\cal L}$ contains a term linear in space-time
derivatives then Eq.(\ref{W2}) may be not valid even for constant
field. This is what happens in TMT where the first term in the
action (\ref{S}) is linear in space-time derivatives of
$A_{\alpha\beta\gamma}$ (when using the definition
(\ref{Aabg}))(see also \footnote{A possibility of a vacuum with
non constant 3-form gauge field has been discussed in Footnote 8
of the Weiberg's review\cite{Weinberg1}}). Then instead of
Eq.(\ref{W2}) which appears to be an identity in this case, the
Euler-Lagrange equations for $A_{\alpha\beta\gamma}$ look
\begin{equation}
\partial_{\mu}\frac{\partial({\Phi L_1})}{\partial
A_{\alpha\beta\gamma,\mu}}=0, \label{W6}
\end{equation}
which are nontrivial even for constant $A_{\alpha\beta\gamma}$ and
resulting in Eq.(\ref{AdL1}). Note that $\Phi L_1$ is a scalar
density and transforms exactly according to Eq.(\ref{W4}).
Therefore generically (i.e. if $A_{\alpha\beta\gamma}$ are not
constant while other fields are constant), the Lagrangian ${\cal
L}$ satisfying (\ref{W4}) can have the following form
\begin{equation}
{\cal L}=c_1\Phi +c_2\sqrt{-g}\label{W7}
\end{equation}
where $c_1$ and $c_2$ are independent of $g_{\mu\nu}$ and $\Phi$.
This is why the equation
\begin{equation}
\partial{\cal L}/\partial\phi=T^{\mu}_{\mu}\sqrt{-g},\label{W8}
\end{equation}
where $T^{\mu}_{\mu}$ is the trace of the energy-momentum tensor,
obtained by Weinberg\cite{Weinberg1} for all constant $g_{\mu\nu}$
and matter fields, is generically no longer valid.

\item
Let us now note that $\zeta =\zeta_{0}(\phi)$,
Eq.(\ref{zeta-without-ferm-delta=0}), becomes singular
\begin{equation}
|\zeta|\approx\frac{2|V_2|}{|V_{1}+
M^{4}e^{-2\alpha\phi/M_{p}}|}\rightarrow\infty \qquad \text{as}
\qquad \phi\rightarrow\phi_{0}. \label{zeta-singular}
\end{equation}
  In this limit the effective potential
 (\ref{Veff1}) (see also
 Eq.(\ref{Veffvac-delta=0})) behaves as
\begin{equation}
V_{eff}\approx\frac{|V_2|}{\zeta^2}. \label{V-singular}
\end{equation}

Thus, disappearance of the cosmological constant occurs in the
regime where $|\zeta|\rightarrow\infty$. In this limit, the
dynamical role of the terms of the Lagrangian $L_2$ (coupled with
the measure $\sqrt{-g}$) in the action (\ref{totaction}) becomes
negligible in comparison with the terms of the Lagrangian $L_1$
(see also the general form of the action (\ref{S})). A particular
realization of this we observe in the behavior of $V_{eff}$,
Eq.(\ref{V-singular}). It is evident that the limit of the TMT
action (\ref{S}) as $|\zeta|\rightarrow\infty$ is opposite to the
conventional field theory (with only measure $\sqrt{-g}$) limit of
the TMT action discussed in the previous subsection. {\bf From the
point of view of TMT, this is the answer to the question why the
old cosmological constant problem cannot be solved (without fine
tuning) in theories with only the measure of integration}
$\sqrt{-g}$ {\bf in the action}.

\item
Recall that one of the basic assumptions of the Weinberg's no-go
theorem is that all fields in the vacuum must be constant. This is
also assumed for the metric tensor components of which  in the
vacuum must be {\bf nonzero} constants. However, this is not the
case in the fundamental TMT action (\ref{totaction}) defined in
the original (non Einstein) frame if we ask what is the metric
tensor $g_{\mu\nu}$ in the $\Lambda =0$ vacuum . To see this let
us note that in the Einstein frame all the terms in the
cosmological equations are regular. This means that the metric
tensor in the Einstein frame $\tilde{g}_{\mu\nu}$ is always well
defined, including the $\Lambda =0$ vacuum state $\phi =\phi_{0}$
where $\zeta$ is infinite. Taking this into account and using the
transformation to the Einstein frame (\ref{ct}) we see that {\bf
all components of the metric in the original frame $g_{\mu\nu}$ go
to zero overall in space-time as $\phi$ approaches the $\Lambda
=0$ vacuum state}:
\begin{equation}
g_{\mu\nu}\sim \frac{1}{\zeta}\sim V_{1}+
M^{4}e^{-2\alpha\phi/M_{p}}\rightarrow 0 \qquad (\mu,\nu =0,1,2,3)
\qquad \text{as} \quad \phi\rightarrow\phi_{0}. \label{g-degen}
\end{equation}
This result shows that the Weinberg's analysis based on the study
of the trace of the energy-momentum tensor misses any sense in the
case $g_{\mu\nu}=0$.

The metric is an attribute of the space-time term. Hence
disappearance of the metric $g_{\mu\nu}$ in the limit
$\phi\rightarrow\phi_{0}$ means that the strict formulation of the
TMT model (\ref{totaction}) with $V_1<0$ and $V_2<0$ requires a
new mathematical basis. A manifold which is not equipped with the
metric (corresponding to the $\Lambda =0$ vacuum state) emerges as
a certain limit of a sequence of space-times. Thus the model under
consideration is in fact formulated not in a space-time manifold
but rather by means of a set of space-time manifolds. A limiting
point of a sequence of space-times is a "vacuum space-time
manifold" (VSTM) one of the differences of which from a regular
space-time is the absence of the metric $g_{\mu\nu}$.

It follows immediately from (\ref{g-degen}) that $\sqrt{-g}$ tends
to zero like
\begin{equation}
\sqrt{-g}\sim \frac{1}{\zeta^2}\sim \left(V_{1}+
M^{4}e^{-2\alpha\phi/M_{p}}\right)^2\rightarrow 0 \quad \text{as}
\quad \phi\rightarrow\phi_{0}. \label{non-degen-tends}
\end{equation}
Then the definition $\zeta\equiv\Phi/\sqrt{-g}$ implies that the
integration measure $\Phi$ also tends to zero but rather like
\begin{equation}
\Phi\sim \frac{1}{\zeta}\sim V_{1}+
M^{4}e^{-2\alpha\phi/M_{p}}\rightarrow 0 \quad \text{as} \quad
\phi\rightarrow\phi_{0}. \label{Phi-tends}
\end{equation}
Thus both the measure $\Phi$ and the measure $\sqrt{-g}$ become
degenerate  in the $\Lambda =0$ vacuum state $\phi =\phi_{0}$.
However $\sqrt{-g}$ tends to zero more rapidly than $\Phi$.

\item
As we have discussed in detail (see Secs.II, IIIB and
Refs.\cite{GK2},\cite{GK3}), with the original set of variables
used in the fundamental TMT action it is very hard or may be even
impossible to display the physical meaning of TMT models. One of
the reasons is that in the framework of the postulated need to use
the Palatini formalism, the original metric $g_{\mu\nu}$ and
connection $\Gamma^{\alpha}_{\mu\nu}$ appearing in the fundamental
TMT action describe a non-Riemannian space-time. The
transformation to the Einstein frame (\ref{ct}) enables to see the
physical meaning of TMT because the space-time becomes Riemannian
in the Einstein frame. Now we see that the transformation to the
Einstein frame (\ref{ct}) plays also the role of a {\bf
regularization of the space-time metric}: the singular behavior of
the transformation (\ref{ct}) as $\phi\approx\phi_{0}$ compensates
the disappearance of the original metric $g_{\mu\nu}$ in the
vacuum $\phi =\phi_{0}$. As a result of this the metric in the
Einstein frame $\tilde{g}_{\mu\nu}$ turns out to be well defined
in all physical states including the $\Lambda =0$ vacuum state.

\end{itemize}

\section{Discussion and Conclusion}

\subsection{Differences of TMT from the standard field theory in
curved space-time} The main  idea of TMT is that the general form
of the action $\int L\sqrt{-g}d^{4}x$ is not enough in order to
account for some of the fundamental problems of particle physics
and cosmology. The key difference of TMT from the conventional
field theory in curved space-time consists in the
hypothesis\cite{GK2}-\cite{GKatz} that in addition to the term in
the action with the volume element $\sqrt{-g}d^{4}x$ there should
be one more term where the volume element is metric independent
but rather it is determined either by four (in the 4-dimensional
space-time) scalar fields $\varphi_{a}$ or by a three index
potential $A_{\alpha\beta\gamma}$,  see
Eqs.(\ref{S})-(\ref{Aabg}). We would like to emphasize that
including in the action of TMT the coupling of the Lagrangian
density $L_{1}$ with the measure $\Phi$, we modify in general both
the gravitational and matter sectors as compared with the standard
field theory in curved space-time. Besides we made two more
assumptions: the measure fields ($\varphi_{a}$ or
$A_{\alpha\beta\gamma}$) appear only in the volume element; one
should proceed in the first order formalism. {\it These
assumptions constitute all the modifications of the general
structure of the theory we have made as compared with the
conventional  field theory where only the measure of integration
$\sqrt{-g}$ is used in the action principle}. In fact, the
Lagrangian densities $L_{1}$ and $L_{2}$ studied in the present
paper, contain  only such
 terms which should be present in a conventional model with
 minimally coupled to gravity scalar field. In particular there is no need
for the non-linear in kinetic energy terms as well as in the
phantom type terms in the fundamental Lagrangian densities $L_{1}$
and $L_{2}$ in order to obtain a super-acceleration phase at the
late time universe.

After making use of the variational principle and formulating the
resulting equations in the Einstein frame, we have seen that the
effective action (\ref{k-eff}) represents a concrete realization
of the $k$-essence\cite{k-essence} obtained from first principles
of TMT without any exotic terms in the Lagrangian densities.

\subsection{Short summary of results}

\subsubsection{The early universe inflation}

As $\delta =0$, the dynamics of $\phi$ can be analyzed by means of
its effective potential (\ref{Veffvac-delta=0}). As $\phi\ll
-M_{p}$ the effective $\phi$ potential has the exponential form
and it is proportional to the integration constant $M^4$. In other
words, the effective potential governing the dynamics of the early
universe results from the spontaneous breakdown of the global
scale invariance (\ref{st}) caused by the intrinsic feature of TMT
(see Eqs. (\ref{varphi}) and (\ref{app1})). We have seen that
independently of the values of the parameters $V_{1}$, $V_{2}$ and
under very general initial conditions, solutions rapidly approach
a regime characterized by a power law inflation. If $\delta\neq
0$, we deal with the intrinsically $k$-essence dynamics. The
numerical solutions in this case have showed that there is no
qualitative difference from the  power law inflation obtained in
the case with $\delta =0$.

\subsubsection{End of inflation}

 In our toy model there
are three regions of the parameters $V_{1}$ and $V_{2}$ and
appropriate three shapes of the effective potentials, Fig.1.
Therefore three different types of scenarios for exit from
inflation can be realized:

a)  $V_{1}<0$ and $V_{2}<0$, \,  Sec.IVB. In this case the power
law inflation ends with damped oscillations of $\phi$ approaching
the point of the phase plane ($\phi =\phi_0, \dot\phi =0$) where
the vacuum energy $V_{eff}^{(0)}(\phi_0)=0$. This occurs without
fine tuning  of the parameters $V_{1}$, $V_{2}$ and the initial
conditions.

b) $V_{1}>0$ and $b_{g}V_{1}>2V_{2}$, \, Sec.IVC. In this case the
power law inflation monotonically transforms to the late time
inflation asymptotically governed by the cosmological constant
$\Lambda_1$.

Qualitatively the same results are also obtained in  the cases a)
and b) if $\delta\neq 0$.

c) $V_{2}<b_{g}V_{1}<2V_{2}$, \, Sec.IVD. In this case the power
law inflation ends without oscillations at the final value
$\phi_{min}$,
 corresponding to the (non zero) minimum of the effective
potential.

The model we have studied in this paper may be extended by
including the Higgs field, as well as gauge fields and fermions.
It turns out that the scalar sector of such an extended model
enables a scenario which resembles a hybrid
inflation\cite{hybrid}. These results will be presented in a
future publication.

\subsubsection{Cosmological constant problems}

1) {\it The old cosmological constant problem}.
 In Sec.IVA
we have seen in details that if $V_1<0$ then, for a broad range of
other parameters, the vacuum energy turns out to be zero without
fine tuning. This effect is a direct consequence of the TMT
structure which yields the following results: a) the effective
scalar sector potential generated in the Einstein frame is
proportional to {\it a perfect square} of two terms; b) one of
those terms is proportional to the integration constant $\pm M^4$
the appearance of which is also the intrinsic feature of TMT. Note
that the spontaneously broken global scale invariance is not
necessary to achieve this effect\cite{GK3}. If such type of the
structure for the scalar field potential in a conventional (non
TMT) model would be chosen "by hand" it would be a sort of fine
tuning.

In Sec.VI we have explained in details how this result avoids the
well known no-go theorem by Weinberg\cite{Weinberg1} stating that
generically in field theory one cannot achieve zero value of the
potential in the minimum without fine tuning. It is interesting
that the resolution of the old CC problem in the context of TMT
happens in the regime where $\zeta\rightarrow\infty$. From the
point of view of TMT, the latter is the answer to the question why
the old cosmological constant problem cannot be solved (without
fine tuning) in theories with only the measure of integration
$\sqrt{-g}$  in the action.

2) {\it The new cosmological constant problem}. Interesting result
following from the general structure of the scale invariant TMT
model with $V_1>0$ is that the cosmological constant $\Lambda$,
Eq.(\ref{lambda}), is a ratio of quantities constructed from
pre-potentials $V_{1}$, $V_{2}$ and the dimensionless parameter
$b_{g}$. Such structure of $\Lambda$ allows to propose two  ways
(see Sec.VI) for resolution of the problem of the smallness of
$\Lambda$ that should be $\Lambda\sim (10^{-3}eV)^{4}$:

\quad a) The first way is a kind of a {\it seesaw}
mechanism\cite{seesaw}. For instance, if  $V_{1}\sim
(10^{3}GeV)^{4}$ and $V_{2} \sim (10^{18}GeV)^{4}$ then
$\Lambda_1\sim (10^{-3}eV)^{4}$.

\quad b) The second way is realized if the dimensionless
parameters $b_{g}$, $b_\phi$ and $V_2/V_1$ of the action
(\ref{totaction}) are huge numbers of the close orders of
magnitude. For example, if $V_{1}\sim (10^{3}GeV)^{4}$ then for
getting $\Lambda\sim (10^{-3}eV)^{4}$ one should assume that
$b_{g}\sim 10^{60}$. Possibility of this idea means that the
resolution of the new cosmological constant problem may have a
certain relation to
 {\it the correspondence principle} between TMT and
conventional field theories (see details in Sec.VIA2).

\subsubsection{Super-acceleration phase of the Universe.
 }
If no fine tuning of the parameters is made in the fundamental
action, namely if $b_g\neq b_{\phi}$, then our TMT model has big
enough regions in the parameter space where the super-acceleration
phase in the late time universe becomes possible. The appropriate
phantom dark energy asymptotically approaches a cosmological
constant.
 However  it is impossible to obtain {\it a pure classical
solution} which connects the early universe power law inflation
with the late time super-acceleration. This problem  is apparently
related with the toy character of the scenario where the role of
the matter creation has been ignored: in TMT the fermionic matter
generically contributes to the constraint equation for the scalar
field $\zeta$ and so can effect the field $\phi$ dynamics as well.

\subsection{What can we expect from quantization}

In this paper we have studied only classical TMT and its possible
effects in the context of cosmology. However quantization of TMT
as well as influence of quantum effects on the processes explored
in this paper may have a crucial role. We summarize here some
ideas and speculations which gives us a hope that quantum effects
can keep the main results of this paper.

 Recall first two fundamental facts of TMT as a classical field
theory: (a) The measure degrees of freedom appear in the equations
of motion only via the scalar $\zeta$, Eq.(\ref{zeta}); (b) The
scalar $\zeta$  is determined (as a function of matter fields, in
our toy model - as a function of $\phi$) by the constraint which
is nothing but a consistency condition of the equations of motion
(see Eqs.(\ref{app1})-(\ref{app3}) in Appendix A and
Eq.(\ref{constraint2})). Therefore the constraint plays a key role
in TMT. Note however that if we were ignore the gravity from the
very beginning in the action (\ref{totaction}) then instead of the
constraint (\ref{app3}) we would obtain Eq.(\ref{app1}) (where one
has to put zero the scalar curvature). In such a case we would
deal with a different theory. This notion shows that the gravity
and matter intertwined in TMT in a much more complicated manner
than in GR. Hence introducing the new measure of integration
$\Phi$ we have to expect that the quantization of TMT may be a
complicated enough problem. Nevertheless we would like here to
point out that in the light of the recently proposed idea of
Ref.\cite{Giddings}, the incorporation of four scalar fields
$\varphi_a$  together with the scalar density $\Phi$,
Eq.(\ref{Phi}), (which in our case are the measure fields and the
new measure of integration respectively), is  a possible way to
define local observables in the local quantum field theory
approach to quantum gravity. We regard this result as an
indication that the effective gravity $+$ matter field theory has
to contain the new measure of integration $\Phi$ as it is in TMT.

 The assumption formulated in item 2 in Sec.IIA, that the
measure fields $\varphi_a$ (or $A_{\alpha\beta\gamma}$) appear in
the action (\ref{S}) {\it only} via the measure of integration
$\Phi$, has a key role in the TMT results and in particular for
the resolution of the old cosmological constant problem. In
principle one can think of breakdown of such a structure by
quantum corrections. However, TMT possesses an infinite
dimensional symmetry mentioned in item 2 of Sec.II which, as we
hope, is able to protect  the postulated structure of the action
from a deformation caused by quantum corrections. Another effect
of quantum corrections is the possible appearance of a nonminimal
coupling of the dilaton field $\phi$ to gravity in the form like
for example $\xi R\phi^2$. Proceeding in the first order formalism
of TMT one can show that the nonminimal coupling can affect the
k-essence dynamics but the mechanism for resolution of the old CC
problem exhibited in this paper remains unchanged. This conclusion
together with expected effect of quantum corrections on the scale
invariance (see our discussion in the paragraph after
Eq.(\ref{Veff=0})) allows us to hope that the exhibited resolution
of the old CC problem holds in the quantized TMT as well.

Quantization of TMT being a constrained system requires developing
the Hamiltonian formulation of TMT. Preliminary consideration
shows that the Einstein frame appears in the canonical formalism
in a very natural manner. A systematic exploration of TMT in the
canonical formalism will be a subject of forthcoming research.

\section{Acknowledgements}

We thank  L. Amendola, S. Ansoldi, R. Barbieri, J. Bekenstein, A.
Buchel, A. Dolgov,  S.B. Giddings, V. Goldstein, P. Gondolo, B-L.
Hu, P.Q. Hung, D. Kazanas, D. Marolf, D.G. McKeon, J. Morris, V.
Miransky,
 H. Nielsen, Y.Jack Ng, H. Nishino, E. Nissimov,  S. Pacheva, L.
Parker, R. Peccei, M. Pietroni, S. Rajpoot, R. Rosenfeld,  V.
Rubakov, E. Spallucci, A. Vikman, A. Vilenkin, S. Wetterich and
A. Zee for helpful conversations on different stages of this
research. Our special gratitude to L. Prigozhin for his help in
implementation of numerical solutions.

\appendix

\section{Equations of motion  in
the original frame}

Variation of the measure fields $\varphi_{a}$ with the condition
$\Phi\neq 0$ leads, as we have already seen in Sec.II, to the
equation $ L_{1}=sM^{4}$ where $L_{1}$ is now defined, according
to  Eq. (\ref{S}), as the part of the integrand of the action
(\ref{totaction}) coupled to the measure $\Phi$.  Equation
(\ref{varphi}) in the context of the model (\ref{totaction}) reads
(with the choice $s=+1$):
\begin{equation}
\left[-\frac{1}{\kappa}R(\Gamma, g)+
\frac{1}{2}g^{\mu\nu}\phi_{,\mu}\phi_{,\nu}\right]e^{\alpha\phi
/M_{p}} -V_{1}e^{2\alpha\phi /M_{p}} =M^{4}, \label{app1}
\end{equation}
It can be noticed that the appearance of a nonzero integration
constant $M^{4}$ spontaneously breaks the scale invariance
(\ref{st}).

Variation of the action (\ref{totaction}) with respect to
$g^{\mu\nu}$ yields
\begin{equation}
-\frac{1}{\kappa}(\zeta +b_{g})R_{\mu\nu}(\Gamma)  +(\zeta
+b_{\phi})\frac{1}{2}\phi_{,\mu}\phi_{,\nu}
+\frac{1}{2}g_{\mu\nu}\left[\frac{b_{g}}{\kappa}R(\Gamma ,g)
-\frac{b_{\phi}}{2}g^{\alpha\beta}\phi_{,\alpha}\phi_{,\beta}
+V_{2}e^{\alpha\phi /M_{p}}\right]
 =0.\label{app2}
\end{equation}

We see that in contrast to field theory models with only the
measure $\sqrt{-g}$, in TMT there are two independent equations
containing curvature. Contracting Eq.(\ref{app2}) with
$g^{\mu\nu}$ and solving Eq.(\ref{app1}) for $R(\Gamma, g)$  we
obtain the following {\it consistency condition} of these two
equations:
\begin{equation}
(\zeta -b_{g})\left(M^{4}e^{-\alpha\phi /M_{p}}+V_{1}e^{\alpha\phi
/M_{p}}\right)+2V_{2}e^{\alpha\phi
/M_{p}}+(b_{g}-b_{\phi})\frac{1}{2}g^{\mu\nu}\phi_{,\mu}\phi_{,\nu}=0,
 \label{app3}
\end{equation}
that we will call {\it the constraint in the original frame}.

It follows from Eqs.(\ref{app1}) and (\ref{app2}) that
\begin{eqnarray}
 \frac{1}{\kappa}R_{\mu\nu}(\Gamma)&=&
 \frac{\zeta +b_{\phi}}{\zeta +b_g}\cdot\frac{1}{2}\phi_{,\mu}\phi_{,\nu}
 \nonumber\\
 &-&\frac{g_{\mu\nu}}{2(\zeta
+b_g)}\left[b_gM^{4}e^{-\alpha\phi /M_{p}}
      +(b_gV_{1}-V_{2})e^{\alpha\phi /M_{p}}
      -(b_g-b_{\phi})\frac{1}{2}g^{\alpha\beta}\phi_{,\alpha}\phi_{,\beta}\right]
\label{app4}
\end{eqnarray}

The scalar field $\phi$ equation of motion in the original frame
can be written in the form
\begin{eqnarray}
&& \frac{1}{\sqrt{-g}}\partial_{\mu}\left[e^{\alpha\phi /M_{p}}
(\zeta +b_{\phi})\sqrt{-g}g^{\mu\nu}\partial_{\nu}\phi\right]
\nonumber\\
&-&\frac{\alpha}{M_{p}}e^{\alpha\phi /M_{p}}\left[(\zeta
+b_g)M^{4}e^{-\alpha\phi /M_{p}}+
[(b_g-\zeta)V_{1}-2V_{2}]e^{\alpha\phi /M_{p}}
-(b_g-b_{\phi})\frac{1}{2}g^{\alpha\beta}\phi_{,\alpha}\phi_{,\beta}\right]
=0 \label{phi-orig}
\end{eqnarray}
where Eq.(\ref{app1})  has been used.

Variation of the action (\ref{totaction}) with respect to the
connection degrees of freedom leads to the equations we have
solved earlier\cite{GK3}. The result is
\begin{equation}
\Gamma^{\lambda}_{\mu\nu}=\{
^{\lambda}_{\mu\nu}\}+\frac{1}{2}(\delta^{\alpha}_{\mu}\sigma,_{\nu}
+\delta^{\alpha}_{\nu}\sigma,_{\mu}-
\sigma,_{\beta}g_{\mu\nu}g^{\alpha\beta}) \label{GAM2}
\end{equation}
where $\{ ^{\lambda}_{\mu\nu}\}$  are the Christoffel's connection
coefficients of the metric $g_{\mu\nu}$ and
$\sigma\equiv\ln\zeta$.

\section{Asymmetry between early and late time dynamics of the
universe as result of asymmetry in the couplings to measures
$\Phi$ and $\sqrt{-g}$ in the action. }

The results obtained in Secs.IV and V depend very much on the
choice of the parameters $V_{1}$, $V_{2}$ and $\delta$ in the
action (\ref{totaction}). Let us recall  that the curvature term
in the action (\ref{totaction}) couples to the measure $\Phi
+b_{g}\sqrt{-g}$ while the $\phi$ kinetic term couples to the
measure $\Phi +b_{\phi}\sqrt{-g}$. This is the reason of
$\delta\neq 0$. If we were choose the fine tuned condition $\delta
=0$ then both the curvature term and the $\phi$ kinetic term would
be coupled to the same measure $\Phi +b_{g}\sqrt{-g}$. One can
also pay attention that depending on the choice of one of the
alternative  conditions $b_{g}V_{1}>2V_{2}$ or $b_{g}V_{1}<2V_{2}$
we realize different shapes of the effective potential if
$b_{g}V_{1}>V_{2}$ (see Fig.1). And again, if instead we were
choose the fine tuned condition $b_{g}V_{1}=V_{2}$ then the action
would contain only one prepotential coupled to the measure $\Phi
+b_{g}\sqrt{-g}$.

So, in order to avoid fine tunings we have  introduced asymmetries
in the couplings of the different terms in the Lagrangian
densities $L_{1}$ and $L_{2}$ to measures $\Phi$ and $\sqrt{-g}$.
In order to display the role of these asymmetries it is useful to
consider what happens if such asymmetries are absent in the action
at all. In other words we want to explore here the gravity+dilaton
model where both $\delta =0$ and $b_{g}V_{1}=V_{2}$. In such a
case the action contains only one Lagrangian density coupled to
the measure $\Phi +b_{g}\sqrt{-g}$:
\begin{equation}
S=\int (\Phi +b_{g}\sqrt{-g}) d^{4}x e^{\alpha\phi /M_{p}}
\left(-\frac{1}{\kappa}R +
\frac{1}{2}g^{\mu\nu}\phi_{,\mu}\phi_{,\nu}-Ve^{\alpha\phi /M_{p}}
\right), \label{S-symm}
\end{equation}
where $V=V_{1}=V_{2}/b_{g}$. An equivalent statement is that
$L_{1}=b_{g}L_{2}$; it is an example of the very special class of
the TMT models where $L_{1}$ is proportional to $L_{2}$.

To see the cosmological dynamics in this model one can use the
results of Sec.IIIB. If we assume in addition $b_{g}>0$ and
$V_{1}>0$, then after the shift $\phi\rightarrow \phi +\Delta\phi$
where $\Delta\phi =-{M_{p}}{2\alpha}\ln(V/M^{4})$ (which is not a
shift symmetry in this case), the effective potential
(\ref{Veffvac-delta=0}) takes the form
\begin{equation}
V_{eff}^{(symm)}(\phi)=\frac{V^{2}}{b_{g}M^{4}}\cosh^{2}(\alpha\phi/M_{p}).
\label{Veff-symm}
\end{equation}
In contrast to general cases ($b_{g}V_{1}\neq V_{2}$)  this
potential has no flat regions and it is symmetric around a certain
point in the $\phi$-axis. This form of the potential (with an
additional  constant) has been used in a model of the early
inflation\cite{Barrow}.

\section{Some remarks on the measure fields independence of
$L_{1}$ and $L_{2}$}

Although we have assumed in the main text that $L_{1}$ and $L_{2}$
are $\varphi_{a}$ independent, a contribution equivalent to the
term $\int f(\Phi/\sqrt{-g})\Phi d^{4}x$ can be effectively
reproduced in the action (\ref{S})  if a nondynamical field
(Lagrange multipliers) is allowed in the action. For this purpose
let us consider the contribution to the action of the form
\begin{equation}
S_{auxiliary}=\int[\sigma\Phi+l(\sigma)\sqrt{-g}]d^{4}x
\label{L-m-action}
\end{equation}
where $\sigma$ is an auxiliary nondynamical field and $l(\sigma)$
is an analytic function. Varying $\sigma$ we obtain
$dl/d\sigma\equiv l^{\prime}(\sigma)=-\Phi/\sqrt{-g}$ that can be
solved for $\sigma$: $\sigma =l^{\prime -1}(-\Phi/\sqrt{-g})$
where $l^{\prime -1}$ is the inverse function of $l^{\prime}$.
Inserting this solution for $\sigma$ back into the action
(\ref{L-m-action}) we obtain
\begin{equation}
S_{aux.integrated}=\int f(\Phi/\sqrt{-g})\Phi d^{4}x
\label{L-m-action-int}
\end{equation}
where the auxiliary field has disappeared and
\begin{equation}
f(\Phi/\sqrt{-g})\equiv l^{\prime -1}(-\Phi/\sqrt{-g})+l(l^{\prime
-1}(-\Phi/\sqrt{-g}))\frac{\sqrt{-g}}{\Phi}.\label{f-aux}
\end{equation}

To see the difference between effect of this type of auxiliary
fields as compared with a model where the $\sigma$ field is
equipped with a kinetic term, let us consider two toy models
including gravity and $\sigma$ field: one - without kinetic term
\begin{equation}
S_{toy}=\int\left[\left(-\frac{1}{\kappa}R+\sigma\right)\Phi
+b\sigma^{2}\sqrt{-g}\right]d^{4}x \label{toy-1}
\end{equation}
and the other - with a kinetic term
\begin{equation}
S_{toy,k}=\int\left[\left(-\frac{1}{\kappa}R+\sigma +
\frac{1}{2}g^{\alpha\beta}\frac{\partial_{\alpha}\sigma
\partial_{\beta}\sigma}{\sigma^{2}}\right)\Phi
+b\sigma^{2}\sqrt{-g}\right]d^{4}x \label{toy-2}
\end{equation}
where $b$ is a real constant. For both of them it is assumed the
use of the first order formalism. The first model is invariant
under local transformations $\Phi\rightarrow J\Phi$, \,
$g_{\mu\nu}\rightarrow Jg_{\mu\nu}$, \, $\sigma\rightarrow
J^{-1}\sigma$ where $J$ is an arbitrary space-time function while
in the second model the same symmetry transformations hold only if
 $J$ is  constant.

Variation of the measure fields $\varphi_{a}$ in the model
(\ref{toy-2}) leads (if $\Phi\neq 0$) to
\begin{equation}
-\frac{1}{\kappa}R+\sigma +
\frac{1}{2}g^{\alpha\beta}\frac{\partial_{\alpha}\sigma
\partial_{\beta}\sigma}{\sigma^{2}}=M^{4},\label{toy-1-phi}
\end{equation}
where $M^{4}$ is the integration constant. On the other hand
varying the action (\ref{toy-2}) with respect to $g^{\mu\nu}$
gives
\begin{equation}
\chi\left(-\frac{1}{\kappa}R_{\mu\nu} +
\frac{1}{2}\frac{\partial_{\mu}\sigma
\partial_{\nu}\sigma}{\sigma^{2}}\right)-
\frac{1}{2}b\sigma^{2}g_{\mu\nu}=0,\label{toy-1-gmunu}
\end{equation}
where $\chi\equiv\frac{\Phi}{\sqrt{-g}}$. The corresponding
equations in the model (\ref{toy-1}) are obtained from
(\ref{toy-1-phi}) and (\ref{toy-1-gmunu}) by omitting the terms
with gradients of $\sigma$. It follows from Eqs.(\ref{toy-1-phi})
and (\ref{toy-1-gmunu}) that
\begin{equation}
\frac{1}{\chi}=\frac{M^{4}-\sigma}{2b\sigma^{2}}
\label{toy-constr}
\end{equation}
This result holds  in both models.

 In the model (\ref{toy-1}), variation of $\sigma$ results in
 $\frac{1}{\chi}=-\frac{1}{2b\sigma}$ which is consistent with
 Eq.(\ref{toy-constr}) only if the integration constant $M=0$.
 This means that through the classical mechanism displayed in TMT,
  it is impossible to achieve
 spontaneous breakdown of the {\it local} scale invariance  the first model
 possesses. This appears consistent with arguments by
 Elitzur\cite{Elitzur} concerning impossibility of a spontaneous
 breaking of a local symmetry without gauge fixing.

 Transition to the Einstein frame where the space-time becomes
 Riemannian is implemented by means of the conformal
 transformation $\tilde{g}_{\mu\nu}=\chi g_{\mu\nu}$. For the
 model (\ref{toy-1}) the gravitational equations in the Einstein frame
 read
\begin{equation}
\frac{1}{\kappa}G_{\mu\nu}(\tilde{g}_{\alpha\beta})=\frac{1}{8b}\tilde{g}_{\mu\nu}.
\label{toy-1-grav}
\end{equation}
This means that the model (\ref{toy-1}) with auxiliary
(nondynamical) field $\sigma$ intrinsically contains a constant
vacuum energy.

In the model (\ref{toy-2}), where $\sigma$ appears as a dynamical
field,  the gravitational equations in the Einstein frame results
from Eq.(\ref{toy-1-gmunu})
\begin{equation}
\frac{1}{\kappa}G_{\mu\nu}(\tilde{g}_{\alpha\beta})=
\frac{(M^{4}-\sigma)^{2}}{8b\sigma^{2}}\tilde{g}_{\mu\nu}
+\frac{1}{2}\left(\frac{\partial_{\mu}\sigma
\partial_{\nu}\sigma}{\sigma^{2}}-
\frac{1}{2}\tilde{g}_{\mu\nu}\tilde{g}^{\alpha\beta}\frac{\partial_{\alpha}\sigma
\partial_{\beta}\sigma}{\sigma^{2}}\right). \label{toy-2-grav}
\end{equation}
It is convenient to rewrite this equation in terms of the scalar
field $\ln\sigma\equiv\phi$:
\begin{equation}
\frac{2}{\kappa}G_{\mu\nu}(\tilde{g}_{\alpha\beta})=V_{eff}(\phi)
\tilde{g}_{\mu\nu} +\frac{1}{2}\left(\partial_{\mu}\phi
\partial_{\nu}\phi-
\frac{1}{2}\tilde{g}_{\mu\nu}\tilde{g}^{\alpha\beta}\partial_{\alpha}\phi
\partial_{\beta}\phi\right), \qquad \text{where} \qquad
V_{eff}(\phi)=\frac{1}{4b\sigma^{2}}(M^{4}e^{-\phi}-1)^{2}.
\label{toy-2-grav-phi}
\end{equation}
 The
$\phi$-equation reads $\Box\phi +V_{eff}^{\prime}(\phi)=0$.
Similar to the general discussion in the main text we see that if
the $\sigma$ field is dynamical then TMT provides the vacuum with
zero energy without fine tuning.

Hence the main difference between the TMT models with auxiliary
and dynamical scalar fields consists in radically different
results concerning the cosmological constant problem.

However it is very unlikely that a nondynamical scalar field will
not acquire a kinetic term after quantum
corrections\cite{Gross-Neveau}. Then it  becomes dynamical which
restores the above results for the model (\ref{toy-2}). This is
why we have ignored the rather formal possibility of introducing
the nondynamical scalars into the fundamental action of the models
studied in this paper.


\begin{thebibliography}{99}
\bibliography{}

\bibitem{Weinberg1}
S. Weinberg, Rev. Mod. Phys. {\bf 61}, 1 (1989).

\bibitem{Weinberg2}
S. Weinberg, astro-ph/0005265;

\bibitem{CC}
S.M.Carroll, Living Rev. Rel. {\bf 4}, 1 (2001); J. Garriga, A.
Vilenkin Phys.Rev.D64:023517,2001.

\bibitem{accel} A. G. Riess et al., {\it Astron. J.} {\bf 116},
1009 (1998); S. Perlmutter {\it et al.},  {\it Astrophys. J.} {\bf
517}, 565 (1999);
 N. Bahcall, J.P. Ostriker, S.J. Perlmutter and
P.J. Steinhardt, {\it Science} {\bf 284}, 1481 (1999); D.N.
Spergel, {\it et al.} [WMAP collaboration], {\it Astrophys. J.
Suppl. 148}, {\bf 175} (2003); A. C. S. Readhead, {\it et al.}
{\it Astrophys. J.} {\bf 609}, 498 (2004);  J. H. Goldstein et
al., {\it Astrophys.J.} 599,  773-785 (2003); R. Rebolo, {\it et
al.}, astro-ph/0402466;  M. Tegmark, {\it et al.} {\it Phys.Rev.}
{\bf D69}, 103501 (2004); E. Hawkins {\it et al.,} {\it
Mon.Not.Roy.Astron.Soc.} {\bf 346}, 78 (2003);
 W. L. Freedman, {\it et al., Astrophys.J.}
{\bf 553},  47 (2001); R. Daly, S.G. Djorgovski, {\it
Astrophys.J.} {\bf 597}, 9, (2003).

\bibitem{coinc}
I. Zlatev, L Wang and P. Steinhardt, {\it Phys. Rev. Lett.} {\bf
82}, 896 (1999).

\bibitem{de-review}
  V. Sahni, A.A. Starobinsky Int.J.Mod.Phys.D9:373-444,2000;
  P.J.E. Peebles and  B. Ratra,
 Rev. Mod. Phys. {\bf 75}, 559 (2003);
T. Padmanabhan, Phys. Rep. {\bf 380}, 235 (2003); N. Straumann,
hep-ph/0604231; T. Padmanabhan, astro-ph/0603114.

\bibitem{Starobinsky}
 U. Alam., V. Sahni and A.A. Starobinsky, {\bf JCAP} 0406:008
 (2004)

\bibitem{Copeland}
E.J. Copeland, M. Sami, S. Tsujikawa, hep-th/0603057.

\bibitem{dm-review}
J.R. Ellis, astro-ph/0304183; J.R. Primack,
Nucl.Phys.Proc.Suppl.{\bf 124}, 3 (2003) ; astro-ph/0205391

\bibitem{quint}
C. Wetterich, Nucl. Phys.  {\bf B302}, 668 (1988). B. Ratra and
P.J.E. Peebles , Phys. Rev. {\bf D37}, 3406 (1988); P.J.E. Peebles
and B. Ratra, Astrophys. J. {\bf 325}, L17 (1988). R. Caldwell, R.
Dave and P. Steinhardt, Phys. Rev. Lett. {\bf 80}, 1582 (1998); N.
Weiss, Phys. Lett. {\bf B197}, 42 (1987); Y. Fujii and T.
Nishioka, Phys. Rev. {\bf D42}, 361 (1990); M.S. Turner and M.
White, Phys. Rev. {\bf D56}, R4439 (1997); P. Ferreira and
M.Joyce, Phys. Rev. Lett. {\bf 79}, 4740 (1997); Phys. Rev. {\bf
D58}, 023503 (1998). E. Copeland, A. Liddle and D. Wands, Phys.
Rev. {D57}, 4686 (1998);
 P. Steinhardt, L Wang and I. Zlatev, Phys. Rev. {\bf
D59}, 123504 (1999).

\bibitem{Amendola}
L. Amendola, Phys. Rev. D{\bf 62}, 043511 (2000); L Amendola and
D. Tocchini-Valentini, Phys. Rev. D {\bf 64}, 043509 (2001)

\bibitem{k-essence}
T. Chiba, T.Okabe and M Yamaguchi, Phys.Rev.D{\bf 62} 023511
(2000); C. Armendariz-Picon, V. Mukhanov and P.J. Steinhardt,
Phys.Rev.Lett. {\bf 85} 4438, 2000; Phys.Rev.D{\bf 63} 103510
(2001); J.K. Erickson, R.R. Caldwell, P.J. Steinhardt, C.
Armendariz-Picon, V. Mukhanov, Phys.Rev.Lett. {\bf 88}121301
(2002); T. Chiba, Phys.Rev.D{\bf 66} 063514 (2002) .

\bibitem{vamp}
J.A.Casas, J. Garcia-Bellido and M. Quiros, {\it Class. Quant.
Grav.} {\bf 9}, 1371 (1992); G.W. Anderson and S.M. Carroll,
astro-ph/9711288; D. Comelli, M. Pietroni and A. Riotto, {\it
Phys. Lett.} {\bf B571}, 115 (2003).


\bibitem{int-q}
L.P. Chimento, A.S. Jakubi and D. Pavon, Phys. Rev. D{\bf 62},
063508 (2000); W. Zimdahl, D. Schwarz, A. Balakin and D. Pavon
Phys. Rev. D{\bf 64}, 063501 (2001); L.P. Chimento,  A.S. Jakubi,
D. Pavon and W. Zimdahl, Phys. Rev. D{\bf 67},083513 (2003); A.A.
Sen and S. Sen, Mod. Phys. Lett. A {\bf 16}, 1303 (2001).
 D.J. Holden and
D. Wands, {\it Phys. Rev.} {\bf D61}, 043506 (2000); A.P. Billyard
and A.A. Coley, {\it Phys. Rev.} {\bf D61}, 083503 (2000); N.
Bartolo and M. Pietroni, {\it Phys. Rev.} {\bf D61}, 023518
(2000);
 M. Gasperini,
{\it Phys. Rev.} {\bf D64}, 043510 (2001); A. Albrecht, C.P.
Burges, F. Ravndal and C. Skordis, astro-ph/0107573; L.P.
Chimento, A.S. Jakubi and D. Pavon, {\it Phys. Rev.} {\bf D62},
063508 (2000); W. Zimdahl, D. Schwarz, A. Balakin and D. Pavon
{\it Phys. Rev.} {\bf D64}, 063501 (2001); L.P. Chimento,  A.S.
Jakubi, D. Pavon and W. Zimdahl, {\it Phys. Rev.} {\bf D67},083513
(2003); A.A. Sen and S. Sen, {\it Mod. Phys. Lett.} {\bf A16},
1303 (2001); G.R. Farrar and P.J.E. Peebles, astro-ph/0307316M;
Axenides, K. Dimopoulos,  {\it JCAP} {\bf 0407}, 010,2004;  M.
Nishiyama, M. Morita, M. Morikawa, astro-ph/0403571; R. Catena, M.
Pietroni and L. Scarabello, {\it Phys.Rev.} {\bf D70}, 103526
(2004).

\bibitem{Chapl}
A.Y. Kamenshchik, U. Moschella and V. Pasquier, Phys. Lett. B{\bf
511}, 265 (2001); N. Bilic, G.B. Tupper and R.D. Viollier, Phys.
Lett. B{\bf 535}, 17 (2002); M.C. Bento, O. Bertolami and A.A.
Sen, Phys. Rev. D{\bf 66} (2002);

\bibitem{phantom}
 R.R. Caldwell, Phys.Lett. B{\bf 545}, 23 (2002); G.W. Gibbons,
hep-th/0302199; P. Singh, M. Sami, N. Dadhich, Phys.Rev. D{\bf 68}
023522 (2003);
 M. Sami, A. Toporensky, Mod.Phys.Lett. A{\bf 19}, 1509 (2004);
 P.F. Gonzalez-Diaz, Phys.Lett. B{\bf 586} 1 (2004); A. Vikman, Phys.Rev.
D{\bf 71}, 023515 (2005);
 P.F. Gonzalez-Diaz, C.L. Siguenza Nucl.Phys. B{\bf697} 363 {2004};
 Wayne Hu,  Phys.Rev. D{\bf 71} 047301 (2005); R.-G. Cai, A.
Wang, JCAP 0503, 002 (2005); S. Nojiri, S. D. Odintsov, S.
Tsujikawa , Phys.Rev. D{\bf 71} 063004 (2005); E. Elizalde, S.
Nojiri, S.D. Odintsov, P. Wang , Phys.Rev. D{\bf71} 103504 (2005);
 H. Stefancic, Phys.Rev. D{\bf 71} 124036 (2005);  L. Amendola, S.
Tsujikawa, M. Sami, Phys.Lett. B{\bf 632}, 155 (2006); I.Ya.
Aref'eva, A.S. Koshelev, S.Yu. Vernov Phys.Rev. D{\bf 72}, 064017
(2005).

\bibitem{tachyon}
G.W. Gibbons, Phys. Lett. B{\bf 537}, 1 (2002); M. Fairbairn and
M.H.G. Tytgat, Phys.Lett.B{\bf 546}, 1 (2002); S. Mukohyama, Phys.
Rev. D{\bf 66}, 024009 (2002);
 A.V. Frolov, L. Kofman, A.A. Starobinsky, Phys.Lett. B{\bf 545}, 8
 (2002); Y.-S. Piao, R.-G. Cai, X. Zhang and T.-Z. Zhang, Phys.
 Rev. D{\bf 66}, 121301 (2002); T. Padmanabhan and T.R. Chounhury,
 Phys. Rev. D{\bf 66}, 081301 (2002); A. Feinstein,
 Phys.Rev. D{\bf 66}, 063511 (2002).

 \bibitem{brane}
 L. Randall and R. Sundrum, Phys. Rev. Lett. {\bf 83}, 4690
 (1999);  M. Gogberashvili, Int.J.Mod.Phys. D{\bf 11}, 635 (2002); T. Shiromizu, K.
Maeda and M. Sasaki, Phys. Rev. D{\bf 62},
 024012 (2000); R. Maartens, D. Wands, B. Bassett and I. Heard,
 Phys. Rev. D{\bf 62} 041301 (2000);  H. Collins and B. Holdom,
 Phys. Rev. D{\bf 62}, 105009 (2000); C. Deffayet, G. Dvali and G.
 Gabadadze, Phys. Rev. D{\bf 65}, 044023 (2002); V. Sahni and Y.
 Shtanov, JCAP {\bf 0311}, 013 (2003).

\bibitem{modified-gravity1} S. Matarrese, C. Baccigalupi and
F. Perrotta, Phys.Rev. D{\bf 70} 061301 (2004)

\bibitem{modified-gravity2}
S. Capozziello, V.F. Cardone, S. Carloni and A. Troisi,
Int.J.Mod.Phys. {\bf D12} 1969 (2003); A. D. Dolgov  and M.
Kawasaki, Phys. Lett {\bf B573}, 1 (2003); S.M.Carroll ,
 A. De Felice , V. Duvvuri ,
  D.A. Easson, M. Trodden and M.S. Turner,
 Phys.Rev. {\bf D71} 063513 (2005).

\bibitem{extra-dim}
M. Pietroni, Phys.Rev. D {\bf 67}, 103523 (2003).

 \bibitem{GK1}
E.I. Guendelman and A.B. Kaganovich, {\it Phys. Rev.} {\bf D53},
7020 (1996); {\it Mod. Phys. Lett.} {\bf A12}, 2421 (1997); {\it
Phys. Rev.} {\bf D55}, 5970 (1997);  {\it Mod. Phys. Lett.} {\bf
A12}, 2421 (1997); {\it Phys. Rev.} {\bf D56}, 3548 (1997); {\it
Mod. Phys. Lett.} {\bf A13}, 1583 (1998).

\bibitem{GK2}
E.I. Guendelman and A.B. Kaganovich, {\it Phys. Rev.} {\bf D57},
7200 (1998).

\bibitem{GK3} E.I. Guendelman and A.B. Kaganovich, {\it Phys.
Rev.} {\bf D60}, 065004 (1999).

\bibitem{G1}
 E.I. Guendelman, {\it Mod.
Phys. Lett.} {\bf A14}, 1043 (1999);
 {\it Class. Quant. Grav.} {\bf 17}, 361 (2000);
gr-qc/9906025; {\it Mod. Phys. Lett.} {\bf A14}, 1397 (1999);
gr-qc/9901067; hep-th/0106085; {\it Found. Phys.} {\bf 31}, 1019
(2001);

\bibitem{K}
A.B. Kaganovich, {\it Phys. Rev.} {\bf D63}, 025022 (2001).

\bibitem{GKatz}
 E.I.
Guendelman and O. Katz, {\it Class. Quant. Grav.}, {\bf 20}, 1715
(2003).

\bibitem{G}
E.I. Guendelman, {\it Phys. Lett.} {\bf B412}, 42 (1997); E.I.
Guendelman, gr-qc/0303048; E.I. Guendelman and E. Spallucci,
hep-th/0311102.

\bibitem{GK5}
E.I. Guendelman and A.B. Kaganovich, {\it Int. J. Mod. Phys.} {\bf
A17}, 417 (2002).


\bibitem{GK6}
E.I. Guendelman and A.B. Kaganovich, {\it Mod. Phys. Lett.} {\bf
A17}, 1227 (2002).

\bibitem{GK7}
E.I. Guendelman, A.B. Kaganovich,  hep-th/0411188;  gr-qc/0603070,
to appear in  Int.J.Mod.Phys.A.

\bibitem{GK8}
E.I. Guendelman, A.B. Kaganovich, hep-th/0603229

\bibitem{hodge}
 H. Nishino, S. Rajpoot, {\it Mod.Phys.Lett}.  A{\bf 21}, 127 (2006).

\bibitem{Mstring}
E.I. Guendelman, {\it Class.Quant.Grav}. {\bf 17}, 3673 (2000);
E.I. Guendelman,  {\it Phys. Rev}. D{\bf 63}, 046006 (2002); E.I.
Guendelman, A.B.  Kaganovich,  E. Nissimov, S. Pacheva,
 {\it Phys. Rev.} D{\bf 66}, 046003 (2002); hep-th/0210062;
hep-th/0304269; E.I. Guendelman, A. Kaganovich, E. Nissimov, S.
Pacheva, {\it Phys.Rev.} D{\bf 72}, 086011 (2005).

\bibitem{Carroll}
S.M. Carroll, {\it Phys. Rev. Lett.} {\bf 81}, 3067 (1998).

\bibitem{power-law}
 L.F. Abbott, M.B. Wise, Nucl.Phys. B{\bf 244}, 541 (1984);
  F. Lucchin, S. Matarrese, Phys.Rev. D{\bf 32}, 1316 (1985);
  J.D. Barrow,
 Phys.Lett. B{\bf 187}, 12 (1987);
J. Yokoyama, Kei-ichi Maeda, Phys.Lett. B{\bf 207}, 31 (1988);
A.R. Liddle, Phys.Lett. B{\bf 220}, 502 (1989);
 V. Muller, H.J. Schmidt, A.A. Starobinsky, Class.Quant.Grav. {\bf 7}, 1163
 (1990); B. Ratra, Phys.Rev. D{\bf 45}, 1913 (1992).

\bibitem{Halliwell} J.J. Halliwell, Phys. Lett.
B{\bf185}, 341 (1987).

\bibitem{Vikman}
A. Vikman, Phys.Rev. D{\bf 71}, 023515 (2005).

\bibitem{seesaw}
 N. Arkani-Hamed, L.J. Hall, C.F. Kolda, H.Murayama, Phys.Rev.Lett. {\bf 85}, 4434
 (2000).

\bibitem{hybrid}
A.D. Linde, Phys.Rev.D49:748-754,1994.

\bibitem{Giddings}
S.B. Giddings, D. Marolf, J.B. Hartle, hep-th/0512200.

\bibitem{Barrow}
J.B. Barrow, Phys. Rev. D{\bf 49}, 3055 (1994).

\bibitem{Elitzur}
S. Elitzur Phys.Rev. D{\bf 12}, 3978 (1975).

\bibitem{Gross-Neveau}
D.J. Gross, A. Neveu, Phys.Rev. D{\bf 10}, 3235 (1974).

\end{thebibliography}
\end{document}